\documentclass[journal=jctcce,layout=twocolumn,manuscript=article]{achemso}
\setkeys{acs}{articletitle=true}
\setkeys{acs}{maxauthors=10}
\setkeys{acs}{etalmode=truncate}

\usepackage[numbers]{natbib}
\usepackage{amsmath}
\usepackage{amsfonts}
\usepackage{amssymb}
\usepackage{xcolor}
\usepackage{mathtools}
\usepackage{braket}
\usepackage{siunitx}
\usepackage{multirow}
\usepackage{longtable}
\usepackage{upgreek}
\usepackage{booktabs}
\usepackage[T1]{fontenc}
\usepackage[utf8]{inputenc}
\usepackage[version = 3]{mhchem}
\usepackage[normalem]{ulem}
\usepackage{cancel}
\definecolor{ao}{rgb}{1.0, 0.13, 0.32}
\definecolor{brightmagenta}{rgb}{0.90, 0.15, 0.55}

\definecolor{rugcolor}{HTML}{D81159}

\definecolor{mycyan}{RGB}{0, 150, 150} % Kolor kresek
\definecolor{headergray}{RGB}{240, 240, 240} % Tło nagłówka

\definecolor{ola}{HTML}{A020F0}

\definecolor{srdftcolor}{HTML}{BC5090}
\definecolor{srpdftcolor}{HTML}{58508D}
\definecolor{pdftcolor}{HTML}{FF6361}
\definecolor{accolor}{HTML}{FCBF49}
\definecolor{srdftmeanbarc}{HTML}{993B73}
\definecolor{srpdftmeanbarc}{HTML}{5B5683}
\definecolor{pdftmeanbarc}{HTML}{FF3A39}
\definecolor{acmeanbarc}{HTML}{A86F00}

\title{
Dynamic electron correlation energy for multireference
wavefunction methods from one- and two-electron reduced density matrices}
\author{Micha{\l} Hapka}
\affiliation{University of Warsaw, Faculty of Chemistry, ul.\ L.\ Pasteura 1, 02-093 Warsaw, Poland}

\author{Aleksandra  Tucholska}
\affiliation{J. Heyrovský Institute of Physical Chemistry, Academy of Sciences of the Czech Republic, 18223 Prague 8 Czech Republic}

\author{Katarzyna Pernal}
\email{pernalk@gmail.com}
\affiliation{Institute of Physics, Lodz University of Technology, ul. Wolczanska 217/221, 93-005 Lodz, Poland}

\begin{document}

\begin{abstract}
Efficiently recovering dynamic correlation in strongly correlated systems without incurring prohibitive computational costs remains a central challenge in quantum chemistry. In this Perspective, we review and benchmark methods capable of recovering dynamic correlation for multireference wave functions exclusively from low-order reduced density matrices and densities. These approaches require at most the two-electron reduced density matrix of the reference wave function and fall into two categories: density functional theory (DFT)-based methods and purely \emph{ab initio} multireference adiabatic connection (AC) methods. The former include MC-srDFT, which recovers dynamic correlation through a short-range exchange--correlation functional depending on the charge and spin densities, 
as well as MC-PDFT and MC-srPDFT, which employ translated functionals that additionally depend on the on-top pair density. Within the post-CASSCF framework, we perform a direct, head-to-head benchmark of these approaches under identical computational settings (including active spaces and basis sets) against challenging multireference problems, including singlet--triplet gaps in organic biradicals, excitation energies, and spin-state splittings in iron complexes. Among the DFT-based methods, MC-srPDFT emerges as the most accurate, underscoring the benefit of incorporating the on-top pair density. However, all considered DFT-based methods fail to provide reliable spin-state energetics for transition-metal complexes. Conversely, linearized AC0 rivals or outperforms more computationally expensive second-order perturbation theory approaches across all benchmark sets. We discuss these findings in the context of alternative formulations and existing literature, highlighting critical limitations and identifying promising directions for the future development of scalable multireference methods.
\end{abstract}

\maketitle

Accurately describing the electronic structure of molecules exhibiting  strong electron correlation or near-degeneracy of electronic energy levels remains one of the central challenges in quantum chemistry. In such cases, a single Slater determinant fails to capture the essential physics and multireference methods become indispensable. Going beyond single-reference description is crucial for understanding the electronic properties of open-shell species, transition metal complexes, diradicals, and systems undergoing bond-breaking or photochemical processes.

Exact electronic wavefunction theory (WFT) methods can, in principle, be systematically improved to reach the  full configuration interaction (FCI) limit. However, in practice, this is only feasible for small molecules due to the exponential growth of computational cost. A common strategy is to approximate the exact wavefunction using a simplified reference that captures the dominant configurations. Among these, the most versatile is the complete active space (CAS) approach, widely used in multireference molecular systems \cite{roos1980complete,roos1987complete,szenes2024striking}. Nevertheless, CAS is typically limited to active spaces of around 20 electrons and orbitals. Developments aimed at approaching the FCI limit, such as selected configuration interaction (sCI) \cite{smith2017cheap,iCISCF,DressedSCI2018}, iterative CI (iCI),\cite{liu2014sds} FCI quantum Monte Carlo (FCIQMC)\cite{booth2009fermion} or density matrix renormalization group (DMRG) \cite{DMRGSCF2017a,DMRGSCF2017b} have extended this limit, but even with these large-active-space methods, achieving quantitative accuracy requires accounting for electron correlation effects beyond the CAS description. 

High computational cost and complexity of wavefunction-based methods have motivated the development of theories that replace the many-electron wavefunction with reduced variables. Density functional theory (DFT) is the most prominent example and remains a central tool in computational chemistry.\cite{teale2022dft} Despite its tremendous success, the Kohn–Sham framework of DFT is inherently single-reference in character, and approximate exchange–correlation functionals often struggle to describe strongly correlated systems and long-range correlation effects. 

Low-order  reduced density matrices contain substantially richer information about electronic structure than the electron density and are therefore better suited for describing strongly correlated systems. For electronic systems, the exact energy can in principle be recovered from the two-electron reduced density matrix (2-RDM).\cite{erdahl2012density} However, variational minimization with respect to the 2-RDM yields the exact ground-state energy only for $N$-representable 2-RDMs. Since the complete set of necessary and sufficient $N$-representability conditions is unknown, practical approaches rely on approximate necessary conditions that are generally insufficient to guarantee accuracy superior to that of WFT methods of comparable cost.\cite{Eugene:24} Consequently, despite recent progress,\cite{mazziotti2023quantum,maradzike2020red,eugene2024variational} highly accurate 2-RDM methods remain largely limited to relatively small ground-state systems. 

Unlike variational 2-RDM methods, one-electron reduced density matrix functional theory (RDMFT)\cite{Gilbert:75} does not suffer from the $N$-representability problem.\cite{Coleman:63} Its main challenge is the approximation of the electron--electron interaction energy functional. Several approximate RDMFT functionals with restricted applicability have nevertheless been developed.\cite{Pernal:16} In particular, Piris' series of functionals (PNOFs), which reconstruct parts of the 2-RDM under approximate $N$-representability constraints,\cite{piris2014perspective,Piris:24} have shown reliable performance for multireference systems while maintaining a favorable balance between accuracy and computational cost.\cite{piris2017global,piris2021global}

Despite recent progress,\cite{piris2024advances,Lemke:22,schouten2025two,erhard2026nec} and many successful applications, approaches relying solely on low-order RDMs appear less reliable than methods that retain a compact, physically meaningful wavefunction reference and recover the remaining, predominantly dynamic, correlation energy through reduced quantities. Multireference methods for dynamic correlation are typically based either on configuration interaction (MRCI) or on perturbation theory (PT). MRCI approaches\cite{siegbahn1980direct,saitow2015fully} are, in principle, systematically improvable; however, this comes at rapidly increasing computational cost, with MRCISD being the practical limit for small- and medium-size systems. Moreover, truncated CI expansions are not size-consistent, which necessitates the use of corrections such as the Davidson correction.\cite{langhoff1974configuration} Noteworthy are recent efforts to reduce the cost of fully internally contracted MRCI methods by avoiding the explicit construction of the four-electron reduced density matrix (4-RDM) through cumulant approximations.\cite{saitow2013multireference}

Multireference perturbation theory (MRPT) remains the most widely used approach for recovering dynamic correlation energy on top of multireference wavefunctions. Its two most popular variants are CASPT2\cite{JPC94,JCP96} and NEVPT2.\cite{nevpt2} While both are second-order methods, extending the perturbation expansion to higher orders leads to substantially increased computational cost without guaranteed improvements in accuracy.\cite{kempfer2025efficient}

CASPT2 employs the one-electron Fock operator as the zeroth-order Hamiltonian which can introduce intruder-state problems. To alleviate this issue, practical implementations typically employ level shifts,\cite{forsberg1997multiconfiguration} while IPEA shifts\cite{ghigo2004modified} are frequently introduced to better balance open-shell and closed-shell states. Although these shifts generally improve results, their optimal use and parameter values remain a subject of ongoing debate.\cite{zobel2017ipea} A more fundamental limitation is that standard CASPT2 implementations require up to four-electron reduced density matrices, whose construction and storage become bottlenecks in calculations with very large active spaces. To reduce this burden, cumulant-expansion approximations have been proposed,\cite{kurashige2014complete} but typically they compromise the accuracy of the method.\cite{phung2016cumulant}

In contrast to CASPT2, NEVPT2 avoids intruder-state problems and the need for empirical shifts by employing the Dyall Hamiltonian\cite{dyall1995choice} as the zeroth-order operator,  which explicitly includes two-electron interactions within the active space. 
However, NEVPT2 suffers  the same scaling limitations as CASPT2: standard formulations demand higher-order, 3- and 4-RDMs.\cite{zgid2009study} Attempts to avoid their explicit construction through cumulant-expansion approximations have  met with limited success, as they may inadvertently reintroduce false intruder states and significantly degrade accuracy.\cite{zgid2009study,guo2021approximations,guo2021approximations2}

Alongside MRPT developments, an alternative class of methods has emerged that formulates dynamic correlation directly in terms of low-order (1- and 2-) RDMs of a multireference wavefunction. These approaches avoid the need for higher-order RDMs entirely thereby circumventing the computational bottleneck of CASPT2 and NEVPT2, while approaching comparable accuracy.

The first group of methods combines WFT and DFT by modifying both the form and the underlying variables of exchange-correlation functionals. 
A rigorous WFT/DFT hybrid was pioneered by Savin and co-workers through a range separation of the electron--electron interaction that avoids the double counting of correlation.\cite{Savin:95,Savin:96} In this range-separated framework, multireference effects are described by the wavefunction, whereas dynamic correlation is accounted for by tailored short-range density functionals.\cite{Toulouse:05,Stoyanova:13} Although the original MC-srDFT formalism was developed for ground states, it has since matured into a general framework\cite{Hedegard:20,pernal2022range} capable of describing excited states,\cite{Toulouse:2004,Fromager:07,Fromager:09,Fromager:10,Hedegard:15,Hubert:16b,hedegaard2017ass,hubert2016inv,kjellgren2019tri} second-order properties including spin splittings\cite{hedegaard2018mul,jorgensen2024perspective} and polarizabilities,\cite{hapka2026opt} and has been extended to embedding schemes.\cite{van2022multiconfigurational}

%Another DFT-based approach considered in this perspective is multiconfiguration pair-density functional theory (MC-PDFT), introduced by Gagliardi and co-workers.\cite{li2014multiconfiguration} In contrast to MC-srDFT, MC-PDFT is based on the idea, originally proposed by Becke and co-workers,\cite{becke1995extension} that the on-top pair density is more sensitive to static correlation effects than electron density and therefore provides a more effective variable for exchange--correlation functionals in multireference situations. In MC-PDFT, conventional density functionals are ``translated'',\cite{li2014multiconfiguration,carlson2015multiconfiguration,rodrigues2023multi} i.e., reformulated in terms of the electron density and on-top pair density,\cite{pdft:annurev} as discussed in the next section. The on-top pair density is obtained from a multireference wave function, most commonly CASSCF. MC-PDFT has been extensively benchmarked and applied to a broad range of strongly correlated problems.\cite{Gagliardi:weak,Gagliardi_benzene,Gagliardi_splittings,wilbraham2017mul,stoneburner2020tra,rodrigues2025magnetic,jorgensen2024perspective,Gagliardi_strongly,king2022lar}

Another DFT-based approach featured in this Perspective is multiconfiguration pair-density functional theory (MC-PDFT), introduced by Gagliardi and co-workers.\cite{li2014multiconfiguration} In contrast to MC-srDFT, MC-PDFT leverages an idea originally proposed by Becke et al.,\cite{becke1995extension} that the on-top pair density is more sensitive to static correlation effects than electron density, making it a more effective variable for exchange--correlation functionals in multireference situations. In MC-PDFT, conventional density functionals are ``translated'',\cite{li2014multiconfiguration,carlson2015multiconfiguration,rodrigues2023multi} meaning they are reformulated in terms of the electron density and on-top pair density,\cite{pdft:annurev} as detailed in the next section. The on-top pair density is obtained from a multireference wave function, most commonly at the CASSCF level. MC-PDFT has been extensively benchmarked and applied to a broad range of strongly correlated problems.\cite{Gagliardi:weak,Gagliardi_benzene,Gagliardi_splittings,wilbraham2017mul,stoneburner2020tra,rodrigues2025magnetic,jorgensen2024perspective,Gagliardi_strongly,king2022lar} Spin–orbit couplings,\cite{zhou2021cal} analytic gradients,\cite{sand2018ana,scott2020ana} and dipole moments\cite{lykhin2021dip,zhou2022ele} are currently available. Multistate MC-PDFT formulations \cite{bao2020mul,bao2020com,hennefarth2023lin} have been demonstrated to recover the correct topology of potential energy surfaces in the vicinity of conical intersections\cite{hennefarth2024sem} and enable transition dipole moment calculations.\cite{lykhin2022dip}. Recently, the method has been extended to the density matrix embedding framework both for solids\cite{mitra2023den} and molecules.\cite{verma2025den}
%
%Moreover, the method has been extended to yield spin–orbit couplings,\cite{zhou2021cal} analytic gradients,\cite{sand2018ana,scott2020ana} and dipole moments\cite{lykhin2021dip,zhou2022ele}, while its multistate formulations \cite{bao2020mul,bao2020com,hennefarth2023lin} have been demonstrated to recover the correct topology of potential energy surfaces in the vicinity of conical intersections\cite{hennefarth2024sem} and enable transition dipole moment calculations.\cite{lykhin2022dip}.

%To further reduce the static-correlation error present in short-range density functional approximations, which, although smaller than in their parent functionals, can still remain significant,\cite{Mussard:17} the idea of translating functional variables, similarly to MC-PDFT, was extended to the short-range framework. This led to the proposal of short-range pair-density functionals (srPDFT).\cite{hapka2020long} By incorporating the on-top pair density, srPDFT improves the description of strongly correlated systems compared to conventional short-range density functionals.\cite{hapka2020long,jorgensen2025multiconfigurational} Promising results have also been reported for excited states and other strongly correlated systems,\cite{helmich2025exc} as discussed in the next section.

The idea of translated functional variables was also adopted within the MC-srDFT framework to alleviate the remaining static-correlation error in srDFT approximations,\cite{Mussard:17} leading to the development of short-range pair-density functional theory (srPDFT).\cite{hapka2020long} By incorporating the on-top pair density, srPDFT  improves the description of strongly correlated systems.\cite{hapka2020long,jorgensen2025multiconfigurational} As discussed in Section~\ref{sec:3}, promising results have also been reported for excited states.\cite{helmich2025exc,rodrigues2026state}

% \kp{Although srDFT approximations exhibit smaller static-correlation error compared to the parent functionals, it can still be substantial.\cite{Mussard:17} To address this, 
% The idea of translating functional variables was recently adapted within MC-srDFT framework\cite{hapka2020long} leading to the introduction of short-range pair-density functionals (srPDFT). This has allowed to reduce the static-correlation error present in short-range density functional approximations.
% By incorporating the on-top pair density, srPDFT improves the description of strongly correlated systems compared to conventional short-range density functionals.\cite{hapka2020long,jorgensen2025multiconfigurational} As detailed in Section~\ref{sec:3}, promising results have been reported for both ground and excited states.\cite{helmich2025exc,rodrigues2026state} }

%The three DFT-based methods discussed above: MC-srDFT, MC-PDFT, and MC-srPDFT, can generally be employed in an \emph{a posteriori} manner, serving as a post-CASSCF \mh{\sout{step following the computation of a} correction based on a previously computed} reference wavefunction and its corresponding density matrices, although MC-srDFT is more commonly used within a self-consistent framework.\cite{hedegaard2018mul}. 
Numerous other hybrid WFT/DFT approaches have been proposed (see, e.g., Ref.~\citenum{ghosh2018combining} for an overview). A comprehensive discussion of all such methods is beyond the scope of this Perspective. The present work focuses primarily on multireference DFT-based approaches  that, in our view, currently offer the most favorable combination of accuracy, versatility, and breadth of applicability.

 A conceptually different class of low-order-RDM-based methods  is provided by multireference adiabatic connection (AC) approximations.\cite{ac_prl} Although originally inspired by the adiabatic connection formalism developed within DFT,\cite{perdew,helgaker_ac,julien_JCTC} these approaches are fully \emph{ab initio} and do not rely on explicit density functional approximations. The AC framework is formally exact,\cite{Pernal:18b} allowing its approximations to be systematically improved. In practical implementations based on extended random phase approximation (ERPA)-type constructions,\cite{erpa1} the required input reduces to the 1- and 2-RDMs of a multireference reference state.\cite{Pastorczak:18a,Pastorczak:18b} The resulting method is generally applicable, size-consistent, and preserves spin symmetry.\cite{tucholska2026mul} 
Driven by successful applications to a range of multireference problems\cite{pastorczak2019capturing,zuzak2024surface,tucholska2026mul,beran2021density,tzeli2024importance} and recent extensions to projection-based embedding schemes,\cite{monino2026projection,monino2026why} multireference AC has emerged as an attractive alternative to MRPT approaches.

The goal of this Perspective is twofold. First, we outline the theoretical foundations and prominent applications of each method. Second, we provide a head-to-head benchmark across a select set of challenging multireference problems. Crucially, all approaches are assessed using strictly uniform, ``standard'' active spaces and basis sets to ensure an unbiased comparison.

The structure of this work is as follows. In Section \ref{sec:2}, we introduce the reduced quantities (densities and reduced density matrices) underlying the low-order RDM methods discussed here and present approaches based on the electron density (srDFT), the on-top pair density (srPDFT and PDFT), and the full 1- and 2-RDMs (AC). In Section \ref{sec:3}, we assess and compare the accuracy of these methods by reviewing previous applications and benchmarking them on selected representative problems. Section \ref{sec:4} provides a summary and a perspective on further developments.

\section{Multireference wavefunction methods and dynamic correlation energy}\label{sec:2}

%The subject of this perspective is electronic structure methods based on a multiconfigurational wavefunction description of electronic states. 
A reference wavefunction, denoted as $\Psi^{\text{ref}}$, provides an approximate description of  the  character of the state of interest. Consequently, the corresponding electronic energy differs from the exact energy, and the difference is generally referred to as the correlation energy:
\begin{equation}
E_{\rm corr} = E_{\rm exact} - E[\Psi^{\text{ref}}]
\label{ecorr}
\end{equation}
Since $\Psi^{\text{ref}}$ is typically constructed to include the principal electronic configurations relevant to the state of interest, part of the total electron correlation, commonly referred to as static correlation, is already captured at the reference level. The remainder, defined in Eq.~\eqref{ecorr}, is referred to as the dynamic correlation energy. The distinction between static and dynamic electron correlation is not sharp, and their numerical values are not uniquely determined because they depend on the particular model used to construct the reference wavefunction. 

The most versatile and widely used strategy for constructing reference wavefunctions suitable for a given chemical or physical problem is based on the concept of an active orbital space. Within this approach, the wavefunction is constructed as a linear combination of single determinants (although it may be advantageous to use configuration state functions) such that a subset of orbitals, the so-called inactive orbitals, remains doubly occupied in every determinant. The active orbitals are those whose occupations vary among the determinants and are used in at least one determinant of the wavefunction. In addition, there exists a set of orbitals that do not enter the CI expansion of $\Psi^{\text{ref}}$; these are referred to as the virtual (or secondary) orbitals.

A convenient and highly successful approach exploiting the concept of active orbitals is the complete active space, introduced by Roos and coworkers~\cite{roos1987complete,roos1980complete}. A CAS wavefunction includes in its CI expansion all possible $N$-electron determinants that can be constructed from a set of $N_{\text{inactive}}$ inactive orbitals and a set of active orbitals describing the remaining $N-2N_{\text{inactive}}$ electrons.

There are three major limitations of the CAS approach: (1) the number of determinants grows exponentially with the number of active orbitals, thereby imposing practical limits on the size of tractable active spaces; (2) there is no universally reliable strategy for selecting the orbitals to be included in the active space; and (3) even when relatively large active spaces are computationally feasible, the resulting wavefunction remains insufficiently correlated, such that the neglected correlation energy may lead to unacceptable errors.

To address the first problem, alternative approaches related to CAS have been developed  to alleviate the exponential growth of the determinant expansion. These methods replace a single large active space with several smaller orbital subspaces and restrict the allowed excitation levels between them. The resulting restricted active space (RAS)\cite{CASSCF-constraint, RASSCF} and generalized active space (GAS)\cite{4C-GASCI2001,GASSCF2011} methods offer considerable flexibility, yet remain highly problem-dependent and require substantial chemical insight to define an appropriate orbital partitioning. Over the last two decades, major progress in the treatment of large active spaces has been achieved by introducing density matrix renormalization group (DMRG) techniques\cite{DMRGSCF2008a,DMRGSCF2009,DMRGSCF2013,DMRGSCF2017a}, into quantum chemistry. As a result, calculations involving dozens of active orbitals, often exceeding 50\cite{cheng2022post}, have become feasible.

These considerations are closely related to the second problem mentioned above, namely the selection of active orbitals. For small active spaces, the prevailing strategy relies on chemical intuition, i.e. choosing orbitals expected to play a major role in the process of interest, such as an electronic transition, a chemical reaction, or intermolecular interactions.\cite{veryazov2011select} For larger active spaces, approaches derived from quantum information theory, exploiting measures such as correlation entropy and mutual information, have proven useful.\cite{stein2016automated} Automated active-space selection schemes have also been proposed, although they may encounter difficulties in maintaining a consistent set of active orbitals along a potential energy surface~\cite{elvira2017aut,bao2018automatic,bensberg2023corresponding}. Orbital entropies can also be approximated from Hartree-Fock quantities~\cite{king2021ranked}, and combined with variational active-space selection~\cite{king2023variational}, though the selected active space is tailored to the method used in the selection step.

A typical multireference computational protocol begins with the construction of a model (reference) wavefunction, $\Psi^{\text{ref}}$. In the second step, auxiliary reduced quantities corresponding to $\Psi^{\text{ref}}$ are computed and used to evaluate the total electronic energy, including both static and dynamic correlation effects. The accuracy of the final result is determined by the theoretical approximations employed, while the computational cost of the ``post-wavefunction'' calculation is governed primarily by the complexity of computing and handling these reduced quantities. Before reviewing methods that rely exclusively on one- and, at most, two-electron reduced quantities, we briefly introduce them in the next section.

\subsection{One- and two-electron reduced density functions}\label{sec:rdms}
Reduced quantities (densities and density matrices) provide a convenient framework for extracting physically relevant information from many-electron wavefunctions while avoiding the explicit manipulation of the full wavefunction.\cite{lowdin1955quantum,davidson2012reduced}
Reduced density matrices (RDMs), arise from partially tracing the information contained in an $N$-electron wavefunction $\Psi$. If only the diagonal  part  of an RDM in the position representation is considered, one obtains the corresponding densities. 

The $k$-particle reduced density matrix ($k$-RDM) in the orbital representation is defined as%
\begin{equation}
\Gamma_{p_{1}p_{2}...p_{k},q_{1}q_{2}...q_{k}}^{(k)}=\left\langle \Psi|\hat
{a}_{q_{1}}^{\dagger}\hat{a}_{q_{2}}^{\dagger}\ldots\hat{a}_{q_{k}}^{\dagger
}\hat{a}_{p_{k}}\ldots\hat{a}_{p_{2}}\hat{a}_{p_{1}}|\Psi\right\rangle
\label{kRDM}
\end{equation}
where the indices of the creation and annihilation operators, $\hat{a}_{q}^\dagger$ and $\hat{a}_{p}$, refer to the chosen orbital basis. For a system containing $N$ electrons, the order of the RDM ranges from $k=1$ to $k=N$. The $N$-particle RDM contains complete information about the wavefunction $\Psi$, while lower-order RDMs provide progressively contracted descriptions.

If the orbital basis contains $M$ orbitals, the storage and manipulation of a $k$-RDM formally scale as $M^{2k}$, making high-order RDMs computationally demanding. Consequently, CAS-based multireference methods designed to recover dynamic correlation become computationally expensive when they involve RDMs of higher order than $k=2$. In particular, starting from $k=3$, the computational cost scales  as the 6th or higher power of the number of active orbitals. One possible strategy to alleviate this difficulty is the cumulant expansion, in which higher-order RDMs are expressed in terms of lower-order RDMs and an additional remainder term, the cumulant matrix~\cite{kutzelnigg1997normal,colmenero1993approximating1,colmenero1993approximating2}. Such approximations have been explored in multireference second-order perturbation theories involving 3- and 4-RDMs. However, the results obtained so far have been unsatisfactory: significant losses of accuracy have been reported, and in some cases spurious intruder states were observed.\cite{zgid2009study,guo2021approximations}

The two-electron reduced density matrix (2-RDM), denoted as $\Gamma$, is obtained from Eq.~(\ref{kRDM}) for $k=2$. When $\Gamma$ corresponds to a CAS reference wavefunction, it is partially factorizable, i.e., it can be expressed as an antisymmetrized product of 1-RDMs, denoted as $\gamma$, unless all orbital indices are active
\begin{equation}
\gamma_{pq}=\left\langle \Psi|\hat{a}_{q}^{\dagger}\hat{a}_{p}|\Psi
\right\rangle \label{1rdm}%
\end{equation}
namely%
\begin{equation}
\Gamma_{pqrs}^{\text{CAS}}=\left\{
\begin{array}
[c]{cc}%
\Gamma_{pqrs}^{\text{act}} & pqrs\in\text{active}\\
\gamma_{pr}\gamma_{qs}-\gamma_{ps}\gamma_{qr} & \text{otherwise}%
\end{array}
\right.
\label{eq:2rdmcas}
\end{equation}
% The elements of 1-RDM are nonzero only when neither of the indices of
% $\gamma_{pq}$ corresponds to a virtual orbital, and in the representation of
% the natural spinorbitals, diagonalising $\gamma$, the diagonal elements of
% $\gamma$ - natural spinorbitals occupation numbers $\left\{  n_{p}\right\}  $,
% are either fractional or integer, $0$ or $1$ if they correspond to a vritual
% or inactive spinorbital%
% \begin{equation}
% n_{p}=\left\{
% \begin{array}
% [c]{cc}%
% 0<n_{p}^{\text{act}}<1 & p\in\text{active}\\
% 1 & p\in\text{inactive}\\
% 0 & p\in\text{virtual}%
% \end{array}
% \right.
% \end{equation}
In the position representation, the 2-RDM becomes a function of four variables, $\Gamma=\Gamma(x_{1},x_{2};x_{1}^{\prime},x_{2}^{\prime})$, where
$x$ denotes combined cartesian and spin coordinates. A further compression of information, still leading to a two-electron
function, is obtained by considering the diagonal part of the 2-RDM, i.e., by
setting $x_{1}=x_{1}^{\prime}$ and $x_{2}=x_{2}^{\prime}$. The resulting
function, $\rho^{(2)}(x_{1},x_{2})=\Gamma(x_{1},x_{2};x_{1},x_{2})$, is the electron pair density and follows from the 2-RDM in the matrix representation as
\begin{equation}
\rho^{(2)}(x_{1},x_{2})=\sum_{pqrs}\Gamma_{pqrs}\ \varphi_{p}(x_{1}%
)\varphi_{q}(x_{2})\varphi_{r}(x_{1})^{\ast}\varphi_{s}(x_{2})^{\ast}%
\end{equation}
Taking the reduction one step beyond the pair density, setting $\mathbf{r}_{1}=\mathbf{r}_{2}$ in the spin-summed pair density yields a function known as the on-top pair density\cite{perdew1997top} (OTPD), $\Pi(\mathbf{r})$%
\begin{equation}
\Pi(\mathbf{r})=\sum_{pqrs}\bar{\Gamma}_{pqrs}\ \varphi_{p}(\mathbf{r}%
)\varphi_{q}(\mathbf{r})\varphi_{r}(\mathbf{r})^{\ast}\varphi_{s}%
(\mathbf{r})^{\ast} \label{OT}%
\end{equation}
where $pqrs$ denote indices of the spatial orbitals $\left\{\varphi
_{p}(\mathbf{r})\right\}$, and $\bar{\Gamma}_{pqrs}$ stands for the
spin-summed 2-RDM. Although the OTPD is significantly more compact than the full 2-RDM, its construction  scales formally as the fourth power of the number of orbitals (or active orbitals in the case of a CAS wavefunction).

In the position representation, $\gamma = \gamma(x,x')$, the diagonal of the 1-RDM yields the electron density, $\rho(x)$:
\begin{equation}
\rho(x)=\sum_{pq}\gamma_{pq}\ \varphi_{p}(x)\varphi_{q}(x)^{\ast}%
\end{equation}
The one-electron spin-free functions related to electron density are charge (C) and spin (S) densities, $\rho_{C}(\mathbf{r})$ and $\rho_{S}(\mathbf{r})$,
determined, respectively, by spin-summed and spin-resolved 1-RDM elements as%
\begin{align}
\rho_{C/S}(\mathbf{r})  &  =\sum_{pq}\gamma_{pq}^{C/S}\ \varphi
_{p}(\mathbf{r})\varphi_{q}(\mathbf{r})^{\ast}\label{rhoCS}\\
\gamma_{pq}^{C}  &  =\sum_{\sigma}\gamma_{p_{\sigma}q_{\sigma}}\\
\gamma_{pq}^{S}  &  =\gamma_{p_{\alpha}q_{\alpha}}-\gamma_{p_{\beta}q_{\beta}}
\label{gS}%
\end{align}
where $\sigma=\alpha$ or $\beta$ pertains to spin symmetry.

%In the following, we provide an overview of theories and computational frameworks that, given a reference multiconfigurational wavefunction, allow the evaluation of the total electronic energy including dynamic correlation, using only the one- or two-electron reduced quantities introduced above. We focus on approaches that are generally applicable to states of arbitrary spin symmetry and whose accuracy has been assessed for representative classes of problems.

\subsection{Dynamic correlation energy from electron density: srDFT}\label{sec:srDFT}
%\ range-separated multiconfigurational DFT (srDF)}

A prominent example of a theory that can, in principle, yield the total electronic energy solely from a reduced one-electron quantity, the electron density, is density functional theory. Within the Kohn–Sham (KS) framework, DFT is essentially a single-determinant theory, and conventional exchange–correlation functionals are known to perform poorly for systems that require more than one determinant. To address this limitation, Savin and collaborators initiated the development of MC range-separated methods,\cite{Stoll:85,Savin:96,Pollet:2002p2162,Toulouse:2004,Leininger:97} MC-srDFT, based on the  separation of the electron--electron Coulomb interaction
operator, $r_{12}^{-1}$, into short- and long-range components, $\hat
{\upsilon}_{ee}^{\rm SR}(r_{12})$ and $\hat{\upsilon}_{ee}^{\rm LR}(r_{12})$,
respectively,
\begin{equation}
\frac{1}{r_{12}} = \hat{\upsilon}_{ee}^{\rm SR}(r_{12}) + \hat{\upsilon}_{ee}^{\rm LR}(r_{12})
\label{RS}
\end{equation}
and assuming wavefunction- and density-functional descriptions of the electron
interaction energy in the long- and short-ranges, respectively. 
%The essential
% conditions for the long-range part is that it is finite at the
% electron-electron coalescence, $r_{12}\rightarrow0$, and reduces to the
% Coulomb interaction, $1/r_{12}$, in the large-separation limit%
% \begin{equation}
% \lim_{r_{12}\rightarrow\infty}\hat{\upsilon}_{ee}^{LR}(r_{12})=\frac{1}%
% {r_{12}}\ \ \ .\label{lr1}%
% \end{equation}

MC-srDFT is a ground state theory, i.e.\ it leads to the exact ground state energy upon minimization of the functional
\begin{equation}
E^{\text{MC-srDFT}}[\Psi] = \left\langle \Psi|\hat{T} + \hat{V}_{ne} + \hat{V}_{ee}^{\rm LR}|\Psi\right\rangle +E^{\rm SR}[\rho_{\Psi}]
\label{MC-DFT}
\end{equation}
where $\hat{T}$ and $\hat{V}_{ne}$ are, respectively, the kinetic energy and
electron-nuclei interaction operators, and $\rho_{\Psi}$ stands for the electronic
density corresponding to a wavefunction $\Psi$. The short-range (SR) density
functional, $E^{\rm SR}[\rho]$, is rigorously defined but it remains unknown in its exact form. To facilitate the construction of  approximations, it is typically split into a Hartree functional, $E_{\rm H}^{\rm SR}[\rho]$
\begin{equation}
E_{\rm H}^{\rm SR}[\rho] = \frac{1}{2}\int\int \rho(\mathbf{r}_{1}) \hat{\upsilon}_{ee}^{\rm SR}(r_{12}) \rho(\mathbf{r}_{2}) \, \mathrm{d}\mathbf{r}_{1} \mathrm{d}\mathbf{r}_{2}\ \ \ 
\label{SRH}
\end{equation}
and the exchange-correlation functional $E_{\rm xc}^{\rm SR}[\rho]$ \cite{gori2006properties,paziani2006local}, such that
\begin{equation}
E^{\rm SR}[\rho] = E_{\rm H}^{\rm SR}[\rho] + E_{\rm xc}^{\rm SR}[\rho]\ \ \  \label{EHxc}
\end{equation}

The most widely used approximations for the short-range exchange-correlation functionals have been developed for the long-range electron interaction described by a parameterized error function\cite{pernal2022range}
\begin{equation}
\hat{\upsilon}_{ee}^{\rm LR}(r_{12}) = \frac{\operatorname{erf}(\mu r_{12})}{r_{12}}\ \ \ \label{erf}
\end{equation}
where $\mu$ is the range separation parameter. These functionals are derived from conventional full-range density functionals, such as LDA~\cite{Toulouse:2004,paziani2006local}, PBE~\cite{goll2005short, goll2006short}, or TPSS~\cite{goll2009development}. Consequently, short-range functionals inherit, to some extent,  the fundamental deficiencies of their full-range counterparts, including fractional-spin and fractional-charge errors. However, when combined with wavefunctions that capture most of the static correlation, these errors are significantly reduced.~\cite{Mussard:17}

Short-range functionals have been combined with various wavefunction theories (see Ref.~\citenum{pernal2022range}   for a comprehensive review) and extended beyond ground states to excited states and molecular properties. Combining an approximate wavefunction theory with DFT via range separation [Eqs.~(\ref{RS}) and (\ref{MC-DFT})] provides an efficient means of recovering  dynamic correlation. The computational cost of evaluating the SR functional is negligible compared with that of the underlying wavefunction calculation. Moreover, because SR functionals depend only on local, one-electron quantities such as the electron density and its derivatives, they do not suffer from the slow basis-set convergence typical of \emph{ab initio} methods.~\cite{giner2018curing,franck2015basis}

The accuracy of energies obtained from MC-srDFT with approximate SR functionals inevitably depends on the range-separation parameter, $\mu$ [see Eq.~\eqref{erf}]. Early studies of small atoms and molecules, as well as dissociation energy curves, suggested a ``universally'' optimal value of $\mu = 0.4$~a.u.~\cite{Fromager:07}. Although it has been shown that other values of $\mu$ may be more appropriate for certain applications,\cite{Fromager:09,hapka2020long} most studies to date continue to employ the ``standard'' value.

As in conventional DFT, a formally exact SR functional should depend only on the charge density, $\rho_C$. However, applications of approximate density functionals to open-shell systems have shown that including a dependence on the spin density, $\rho_S$ [cf. Eq.~\eqref{gS}], in the SR exchange-correlation energy
\begin{equation}
E_{\rm xc}^{\rm SR,spin} = E_{\rm xc}^{\rm SR}[\rho_C, \rho_S]
\end{equation}
can be beneficial.\cite{hedegaard2018mul} Several approximations, including the widely used srPBE~\cite{goll2006short} and srLDA\cite{paziani2006local}, have been extended to incorporate spin-density dependence. 
%\kp{\sout{While this improves the description of systems such as dioxygen and iron–water complexes,\cite{hedegaard2018mul} it should not be assumed that spin-dependent SR approximations always outperform their spin-independent variants for open-shell states; see the next section for further discussion.}}

The wavefunction $\Psi$ minimizing the exact functional in Eq.~\eqref{MC-DFT} differs from the true wavefunction; for instance, it is always free of the electron coalescence cusp.\cite{franck2015basis} By definition, the number of significant configurations in $\Psi$ is smaller than in a full CI expansion, justifying shorter CI expansions.\cite{giner2018curing} In MC-srDFT, the minimizing wavefunction can, in principle, be obtained self-consistently by diagonalizing the Hamiltonian containing the LR electron interaction and the SR potential.  However, given the similarity between CAS wavefunctions obtained with the LR or full-range Coulomb operator, and the fact that only approximate SR functionals are available, computing the CAS wavefunction with either Hamiltonian yields essentially equivalent MC-srDFT energies. In practice, a post-CASSCF scheme, in which the CASSCF wavefunction is found using the full-range Hamiltonian and the MC-srDFT energy is evaluated afterward via Eq.~\eqref{MC-DFT}, has been shown to be equally valid and comparably reliable~\cite{hapka2020long}, cf.\ also next section.

The post-CASSCF energy is state-specific, with $\Psi^{\text{CAS}}$ obtained from either state-specific or state-averaged (SA) CASSCF. Variational SA MC-srDFT is more challenging because the effective Hamiltonian differs for each state, namely the SR potential depends on the density of the particular state. This problem can be addressed by using the ensemble-averaged density in the SR potential, shared across all states in the ensemble.\cite{helmich2025exc}

Range-separated methods, which rely on short wavefunction expansions such as CAS,  miss the long-range correlation energy.~\cite{hapka2020much} As a result, the dispersion component of the interaction energy is absent, which severely affects the description of  weakly interacting systems. This limitation has been addressed by introducing long-range correlation corrections, either via second-order perturbation theory~\cite{Fromager:10} (which requires RDMs beyond second order) or through MC adiabatic connection theory~\cite{hapka2020long}, the latter relying solely on the 1- and 2-RDMs.

%Hereafter, we refer to the range-separated energy expression in Eq.~\eqref{MC-DFT} simply as ``srDFT'' to indicate that the dynamic correlation energy is captured by the SR density functional.

\subsection{Dynamic correlation energy from on-top pair
density: PDFT and srPDFT}\label{sec:pdft}

The on-top pair density, defined in Eq.~\eqref{OT}, is a local function of position that describes the probability of finding two electrons at coalescence. Unlike the one-electron density, the OTPD is a two-electron quantity and  conveys information about electron correlation.\cite{carlson2017top,Gritsenko:18} The use of OTPD to improve the description of electron correlation dates back to the work of Perdew, Savin, and Burke~\cite{perdew1995escaping}, who suggested that incorporating OTPD dependence in density functionals could mitigate the spin-symmetry breaking problem of local and GGA approximations. Later, this idea was extended to multiconfigurational reference states, where the local spin-density functional was generalized using the OTPD.\cite{becke1995extension}

Inspired by these works, Gagliardi, Truhlar, and co-workers\cite{Gagliardi:14} proposed an ansatz for the electronic energy that accounts for both the multireference character of the state and the dynamic correlation energy by means of the OTPD functional. The resulting theoretical framework, termed Multiconfigurational Pair-Density Functional Theory (MC-PDFT),\cite{Gagliardi_strongly,pdft:annurev} is based on partitioning the total energy into the sum of the kinetic energy, the electron–nuclear interaction energy, and the Hartree energy, all evaluated with the reference 1-RDM, and the remaining part of the electron–electron interaction energy, the exchange–correlation energy, treated using conventional spin-density functional approximations with modified arguments:
\begin{equation}
\begin{split}
E^{\text{MC-PDFT}}[\Psi^{\text{ref}}] &= 
\left\langle \Psi^{\text{ref}}|\hat{T}+\hat{V}_{ne}|\Psi^{\text{ref}}\right\rangle \\
&+ E_{\rm H}[\rho^{\text{ref}}] + E_{\rm xc}[\tilde{\rho}_{C},\tilde{\rho}_{S}]
\end{split}
\label{PDFT}
\end{equation}
where the Hartree term involves full-range Coulomb operator
\begin{equation}
E_{\rm H}[\rho^{\text{ref}}] = \frac{1}{2}\int\int
\frac{\rho^{\text{ref}}(\mathbf{r}_{1})\rho^{\text{ref}}(\mathbf{r}_{2})}{r_{12}} \, \mathrm{d}\mathbf{r}_{1} \mathrm{d}\mathbf{r}_{2}\ \ \ \label{EH}
\end{equation}
%and is computed for electron density obtained from a given reference wavefunction.

The central idea of MC-PDFT lies in the introduction of auxiliary local functions, $\tilde{\rho}_{C}$ and $\tilde{\rho}_{S}$, constructed from fictitious spin densities, $\tilde{\rho}_{\alpha}$ and $\tilde{\rho}_{\beta}$. These are defined in terms of a reference electron density, $\rho^{\text{ref}}$, scaled by a factor modulated by the ratio of the multiconfigurational OTPD, $\Pi^{\text{ref}}$, and the square of electron density, namely%
\begin{equation}
\tilde{\rho}_{\alpha/\beta}(\mathbf{r})=\frac{\rho^{\text{ref}}(\mathbf{r}%
)}{2}\left(  1\pm\sqrt{1-\frac{2\Pi^{\text{ref}}(\mathbf{r})}{\rho
^{\text{ref}}(\mathbf{r})^{2}}}\right)  \label{translation}%
\end{equation}
and
\begin{align}
\tilde{\rho}_{C} &  =\tilde{\rho}_{\alpha}+\tilde{\rho}_{\beta}\\
\tilde{\rho}_{S} &  =\tilde{\rho}_{\alpha}-\tilde{\rho}_{\beta}%
\label{tilded}
\end{align}

An immediate advantage of MC-PDFT is that it combines wavefunction theory with DFT in such a way that information about static correlation, contained in a multireference wavefunction of the correct spin and spatial symmetry, is transferred to the exchange-correlation functional via the OTPD. In this manner, the typically correct behavior of exchange-correlation functionals for spin-symmetry-broken  solutions is effectively exploited without actually breaking the symmetry. This can be illustrated by considering the dissociation of the H$_{2}$ molecule described with a CAS(2,2) reference wavefunction. Near the equilibrium geometry, where the CAS(2,2) energy lacks dynamic correlation, the wavefunction is dominated by a single determinant. Consequently, $\Pi^{\text{ref}}(\mathbf{r}) \approx \rho^{\text{ref}}(\mathbf{r})^{2}/2$, and $\tilde{\rho}_{\alpha}(\mathbf{r}) = \tilde{\rho}_{\beta}(\mathbf{r}) = \rho^{\text{ref}}(\mathbf{r})/2$. 
As a result, the MC-PDFT energy is effectively equal to that of the chosen density functional approximation evaluated with a spin-unpolarized density, yielding a lower (i.e., more correlated) energy than CAS(2,2).
In the dissociation limit, the OTPD corresponding to CAS(2,2) vanishes~\cite{carlson2017top,Gritsenko:18}. As a consequence, $\tilde{\rho}_{\alpha}(\mathbf{r}) = \rho^{\text{ref}}(\mathbf{r})$ and $\tilde{\rho}_{\beta}(\mathbf{r}) = 0$. In this regime, the exchange-correlation functional $E_{xc}$ is evaluated with  variables that mimic a  spin-symmetry-broken solution. Thus, the entire MC-PDFT dissociation curve of H$_{2}$ is obtained accurately without breaking spin symmetry~\cite{Gagliardi:14}.

The ``translation'' of the electron density and OTPD into auxiliary densities, as shown in Eq.~\eqref{translation}, is the most commonly used approach. However, alternative formulations have also been proposed, differing primarily in the treatment of cases where the argument of the square root becomes negative. While the standard scheme simply sets this argument to zero, complex-valued extensions have recently been explored to handle these cases.\cite{rodrigues2023multi}
Moreover, the original formulation does not include the gradient of the on-top pair density, whereas the ``full'' translation explicitly incorporates this dependence.\cite{carlson2015multiconfiguration}

Originally, and in most common applications, PDFT is used in a post-CASSCF fashion, i.e., the usual CASSCF (or other multiconfigurational wavefunction) calculation is carried out, and the total energy is obtained from Eq.~\eqref{PDFT} in the subsequent step. Recently, self-consistent PDFT\ computational schemes have  also been developed.\cite{scott2024variational,helmich2025exc} 
%\mh{\textit{cite also Helmich-Paris \citenum{helmich2025exc}?}}
%Variational PDFT energy formulation does not bring any
%obvious benefits for the accuracy but opens the doors to geometry
%optimization. 

Unlike MC-srDFT, MC-PDFT is not based on a formally exact framework.
%, and an exact construction of the theory is not known. 
%As a result, several sources of error may arise, including possible correlation double counting, which is expected to become significant mainly for large active spaces~\cite{citepostCAS}. In addition, the accuracy of MC-PDFT is limited by the deficiencies of the underlying exchange–correlation functional, such as self-interaction error~\cite{ourpaper}.
%Despite these limitations, 
In spite of this potential limitation, MC-PDFT has been successfully applied to a wide range of systems, including molecules in both ground and excited states with complex electronic structures,\cite{pdft:annurev,sharma2018active,Gagliardi_benzene,Gagliardi_splittings,bao2018automatic,wilbraham2017mul,stoneburner2018mcp,hoyer2016mul,hennefarth2023lin,ghosh2015mul,wardzala2026multireference} The substantial body of applications provides practical guidance on its reliability and delineates the regimes in which the method can be expected to perform well.

The successful strategy of using the OTPD obtained from multiconfigurational calculations to construct auxiliary quantities that mimic spin-symmetry breaking\cite{Gritsenko:18,hapka2020local,pernal2020embracing} has recently been adopted within the MC-srDFT framework.\cite{hapka2020long,jorgensen2025multiconfigurational} The artificial spin polarization introduced by employing the auxiliary densities, Eq.~\eqref{translation}, in the SR exchange-correlation functional plays a role analogous to replacing a spin-restricted solution with a spin-broken one. As a result, it leads to a reduction of both self-interaction and static-correlation errors.\cite{hapka2020long}
%However, if the OTPD corresponds to a multiconfigurational wavefunction, than
%the fictitious spin polarization is large in the regions of space where the
%static correlation prevails while it remains small in regions dominated by the
%dynamic correlation cite 36 from 07. This has been derived from the behavior
%of the ratio $2\Pi^{\text{ref}}(\mathbf{r})/\rho^{\text{ref}}(\mathbf{r})^{2}$
%the value of which decreases when static correlation effects become important. 

Therefore, using the OTPD in SR density functionals, via the transformation given in Eq.~(\ref{translation}),  provides an alernative, physically motivated way to recover dynamic correlation for a multiconfigurational wavefunction, such as CAS. The total energy is then given by
\begin{equation}
\begin{split}
E^{\text{MC-srPDFT}}[\Psi^{\text{CAS}}] &= \left\langle \Psi^{\text{CAS}}
|\hat{T}+\hat{V}_{ne}+\hat{V}_{ee}^{\rm LR}|\Psi^{\text{CAS}}\right\rangle \\
 & + E_{\rm H}^{\rm SR}[\rho] + E_{\rm xc}^{\rm SR}[\tilde{\rho}_{C},\tilde{\rho}_{S}]
\end{split}
\label{CASsrPDFT}
\end{equation}
where $E_{\rm H}^{\rm SR}[\rho]$ is the SR Hartree energy defined in
Eq.~\eqref{SRH}, while the tilded quantities in the SR exchange-correlation functional are defined in Eqs.~(\ref{translation})-(\ref{tilded}).

Comparing the MC-srPDFT and MC-PDFT energy expressions, Eq.~(\ref{CASsrPDFT}) and Eq.~(\ref{PDFT}), respectively, one observes that in MC-PDFT the exchange–correlation functional is responsible for recovering the full nonclassical part of the electron–electron interaction, whereas in MC-srPDFT it accounts only for the short-range component, typically associated with dynamic correlation. This distinction has important consequences, particularly for the description of molecular interactions. It is well known that a CAS wavefunction alone provides a poor description of dispersion, since the interaction energy lacks long-range correlation contributions~\cite{hapka2020much}. Therefore, MC-srPDFT is expected to benefit significantly from an enlarged active space or even from replacing CAS with a CI wavefunction. The impact on MC-PDFT is less straightforward. Most likely, MC-PDFT applied to molecular interactions would exhibit performance comparable to conventional DFT methods.

Note that MC-PDFT requires only Coulomb two-electron integrals, together with the 1-RDM and OTPD, and is therefore slightly less demanding computationally than MC-srPDFT, which additionally requires modified two-electron integrals. Otherwise, when both methods are used in a post-CASSCF fashion, their overall computational cost is essentially the same.

%Applications so far have been scarce and they showed potential usefulness of
%srPDFT\ for ground and excited states, but poor accuracy has been recently
%reported for transition metals molecules. The discussion is postponed to the
%next section. 

\subsection{Correlation energy from 1- and 2-RDMs: AC0}\label{sec:AC}
The Adiabatic Connection theory provides a general, in principle
exact,\cite{janos,ac_prl,Pernal:18b} framework for calculating the correlation energy $E_{\text{corr}}$, defined as in Eq.~\eqref{ecorr}. It is based on the assumption that there
exists a Hamiltonian $\hat{H}^{(0)}$ such that a given reference wavefunction $\Psi^{\text{ref}}$ is one of its eigenfunctions, $\hat{H}^{(0)}|\Psi^{\text{ref}}\rangle=E^{(0)}|\Psi^{\text{ref}}\rangle$. The AC formalism is established by introducing the adiabatic connection Hamiltonian, $\hat{H}^{\alpha}$, which smoothly interpolates between $\hat{H}^{(0)}$
and $\hat{H}$, recovered at $\alpha=0$ and $\alpha=1$, respectively 
\begin{equation}
    \hat H^{\alpha} = \hat H^{(0)} +\alpha \hat H^{\prime} 
\end{equation}
The correlation energy defined in Eq.~\eqref{ecorr} follows from integrating the perturbed energy along the adiabatic connection path:\cite{Pernal:18b}
\begin{equation}
E_{\text{corr}}^{\text{AC}}=\int_{0}^{1}W(\alpha)\ \text{d}\alpha
\ \ \label{ACcorr}%
\end{equation}
where formally $W(\alpha)=\langle\Psi^{\alpha}|\hat{H}^{\prime}|\Psi^{\alpha
}\rangle-\langle\Psi^{\text{ref}}|\hat{H}^{\prime}|\Psi^{\text{ref}}\rangle$,
and $\Psi^{\alpha}$ is an eigenstate of $\hat H^{\alpha}$  corresponding to the state of interest, which at
$\alpha=0$ coincides with $\Psi^{\text{ref}}$. By applying the anticommutation rules for fermionic creation and annihilation operators, the integrand $W(\alpha)$ can be expressed in terms of one-electron reduced density matrices: 1-RDM and transition reduced density matrices (1-TRDMs)
\begin{equation}
\forall_{\nu\in{\mathcal{H}}_{N}^{\alpha}}\ \ \ \left[  \gamma_{\text{ph}%
}^{\nu}\right]  _{pq}^{\alpha}=\left\langle \Psi^{\alpha}|\hat{a}_{q}%
^{\dagger}\hat{a}_{p}|\Psi_{\nu}^{\alpha}\right\rangle \label{phTRDM}%
\end{equation}
or
\begin{equation}
\forall_{\nu\in{\mathcal{H}}_{N+2}^{\alpha}}\ \ \ \left[  \gamma_{\text{pp}}^{\nu}\right]_{pq}^{\alpha} = \left\langle \Psi^{\alpha}|\hat{a}_{p}\hat{a}_q|\Psi_\nu^\alpha \right\rangle \label{ppTRDM}
\end{equation}
in the particle-hole (ph)\cite{mclachlan,ac_prl} or particle-particle (pp)\cite{van2013exchange,van2014exchange,tucholska2024dua,tucholska2026mul} formulations (ph- and pp-TRDMs are transition density matrices connecting the $N$-electron reference state with excited states containing $N$- and $N+2$-electrons, respectively\cite{tucholska2024dua}).

In the multireference AC methods,  the 1-RDM is assumed to remain frozen along the AC path,\cite{ac_prl,Pastorczak:18a,Pastorczak:18b,pastorczak2019capturing,tucholska2026mul} and the problem of determining the correlation
energy reduces to constructing efficient approximations for 1-TRDMs. For that purpose, the extended random phase approximation\cite{erpa1,rpa_tutorial} originating from equations of motion (EOM) formalism of Rowe,\cite{rowe} has been employed. Within ERPA, 1-TRDMs follow from solving a generalized eigenproblem for the Hessian matrix defined  with a double commutator
\begin{equation}
\left[ \mathbf{A}_{\text{ff}}^\alpha \right]_{IJ} = \left\langle
\Psi^{\text{ref}} \right\vert \left[ \hat{o}_I^{\text{ff}},[\hat{H}^{\alpha}, (\hat{o}_J^{\text{ff}})^{\dagger}]\right]  \left\vert \Psi^{\text{ref}} \right\rangle 
\label{Aff}
\end{equation}
where the operators $\hat{o}_{I}^{\text{ff}}$ are either of the ph or pp types\cite{tucholska2024dua,tucholska2026mul}
\begin{equation}
\forall_{I=pq}\ \ \ \hat{o}_{I}^{\text{ff}}=\left\{
\begin{array}
[c]{cc}%
\hat{a}_{q}^{\dagger}\hat{a}_{p} & \ \ \ \text{ff}=\text{ph}\\
\hat{a}_{p}\hat{a}_{q} & \ \ \ \text{ff}=\text{pp}%
\end{array}
\right.  \ \ \ \label{oI}%
\end{equation}

Expanding the AC integrand, $W(\alpha)$, in the AC correlation energy expression, Eq.~\eqref{ACcorr}, up to the linear term in $\alpha$, leads to the ``linearized AC'' approximation, denoted
AC0 \cite{Pastorczak:18a, Pernal:18b, beran2021density, guo2024spinless},
\begin{equation}
E_{\text{corr}}^{\text{AC0}}=\frac{1}{2}W^{(1)}\label{WAl0}%
\end{equation}
The AC0 approximation can equivalently be derived from the second-order perturbation-theory expression for the correlation energy,\cite{tucholska2024dua}
$E_{0}^{(2)}=\left\langle \Psi_{0}^{(0)}|\hat{H}^{\prime}|\Psi_{0}%
^{(1)}\right\rangle =E_{\text{corr}}^{\text{AC0}}$.
When the reference wavefunction is an HF determinant, AC0
reduces to the MP2 correlation-energy expression. Both AC and AC0 correlation energies vanish in the FCI limit, where the reference wavefunction becomes exact. Moreover,  AC and its approximate variants are size-consistent, a desired property that is not trivially satisfied in multireference dynamic-correlation theories.

Combining AC with ERPA recovers dynamic correlation using only the 1- and 2-RDMs of the reference wavefunction, in addition to one- and two-electron integrals. In particular, for a CAS reference wavefunction, the 2-RDM is partially factorizable, see Eq.~\eqref{eq:2rdmcas}, so that  the AC correlation energy depends only on the full 1-RDM, $\gamma^{\text{CAS}}$, and the active-orbital block of the 2-RDM
\begin{equation}
E_{\text{corr}}^{\text{AC}}[\Psi^{\text{ref}}]=E_{\text{corr}}^{\text{AC}%
}[\gamma^{\text{CAS}},\ \Gamma^{\text{act}}]
\end{equation}

The crucial advantage of the AC0 approximation over AC is that the ERPA equation needs to be solved only for the coupling constant $\alpha=0$. In this limit, the ERPA matrix $\mathbf{A}$ becomes block diagonal\cite{Pastorczak:18a,guo2024spinless,tucholska2026mul} and the dimension of the largest block grows quadratically with the number of active orbitals. As a result, the computational cost of AC0 scales as the 5th power of the number of electrons and the 6th power of the number of active orbitals. It has also been shown that AC can achieve the same formal scaling, albeit with a larger prefactor. This is possible by employing the Cholesky decomposition of Coulomb integrals within the iterative AC scheme, in which higher-order contributions in $\alpha$ are recovered successively at each iteration.\cite{drwal2022efficient} Recently, the favorable scaling of AC0 enabled its successful application to long polyacenes, where the largest system considered involved more than 50 active orbitals.\cite{zuzak2024surface}

The ph formulation of AC0 provides satisfactory accuracy for most applications, with the notable exception of singlet excitations, for which significantly larger errors are observed.\cite{drwal2021exc} 
Drwal et al.\cite{drwal2021exc} demonstrated that these inaccuracies originate from the limitations of the phERPA model which does not recover the negative-transition part of the density linear response function. In other words, for excited-state wave functions, AC based on phERPA lacks negative transitions in the active--active block of the excitation spectrum.\cite{tucholska2024dua} This deficiency can be remedied by explicitly recovering the missing contributions through the solution of phERPA equations for all states lower in energy than the reference wavefunction. This leads to a so-called deexcitation correction to the correlation energy of a given excited state of interest, thereby improving the overall accuracy.\cite{drwal2021exc,guo2024spinless}

A more elegant and versatile solution was later proposed by Tucholska et al.\cite{tucholska2024dua,tucholska2026mul} through a physically motivated combination of the ph and pp variants of AC.
%, leading to the ffAC0 method.\cite{tucholska2024dua} 
%Recently, Tucholska and Pernal\cite{tucholska2024dua,tucholska2026mul} introduced a multireference AC formalism based on the particle-particle decomposition of the 2-RDM. 
The practical implementation of ppAC relies on the solution of ppERPA equations, which reduce in the single-reference limit to ppRPA, originally developed by Yang and co-workers.\cite{van2013exchange,yang2013dou,yang2013ben} The ppAC method can be rigorously combined with phAC by exploiting the duality between the corresponding correlation energy amplitudes.\cite{tucholska2024dua} This strategy is motivated by the identification of the amplitude components underlying the deficiencies of phAC0 for excited states.\cite{guo2024spinless} 
The resulting ffAC0 approach is currently regarded as the most broadly applicable and accurate AC variant for  systems requiring a multireference treatment.\cite{tucholska2026mul}

\section{Performance across representative multireference problems}\label{sec:3}
%
% srPBE spin-functional used only for Fe complexes; in other cases it is srPBE depending on charge density (polarization = 0)
% ffAC0 for Thiel and biradicals; phAC0 for Fe complexes
%
The methods presented in Section \ref{sec:2} recover dynamic correlation energy from one- and two-electron reduced quantities obtained from the underlying multiconfigurational wavefunction, typically CASSCF. They are  applicable to electronic states of arbitrary spatial and spin symmetry, with the latter imposed at the wavefunction level. The performance of these approaches has already been tested for a broad range of chemical problems. To directly compare their accuracy, we consider here three representative classes of multireference problems: singlet--triplet gaps of organic biradicals, excitation energies of organic chromophores, and spin-state splittings in transition-metal complexes. For all methods, we employ computational protocols established in the literature, including consistent choices of active spaces in the CASSCF calculations and basis sets appropriate for the systems under investigation.

All calculations are performed in a post-CASSCF fashion:  a CASSCF calculation
%, either state-specific or state-averaged, 
is followed by construction of the reduced quantities for the state of interest and evaluation of the total energy. The MC-srDFT energy, referred to below as ``srDFT'', is computed according to Eqs.~\eqref{MC-DFT}-\eqref{EHxc} using the short-range PBE exchange-correlation functional.\cite{goll2005short}
MC-PDFT and MC-srPDFT energies, denoted below as ``PDFT'' and ``srPDFT'', respectively, are evaluated from the expressions given in Eqs.~\eqref{PDFT} and \eqref{CASsrPDFT}. The srPDFT calculations employ the short-range PBE functional\cite{goll2005short} with translated densities, see Eq.~\eqref{translation}. The PDFT calculations are based on the same translation and use the PBE functional\cite{perdew1996generalized}.
%, see Eqs.~\eqref{PDFT}. 
A standard value of the range-separation parameter, $\mu=0.4$ a.u., is employed in all range-separated integrals and functionals. 
Finally, the CAS-AC0 energy, denoted ``AC0'', is computed as described in Sec.~\ref{sec:AC}.

%\ola{\sout{For CASSCF calculations, the \textsc{Molpro} \mh{\textit{I think PySCF and Dalton in more cases?}} suite of programs~\cite{Molpro:12} was used, while the subsequent post-CASSCF energy calculations with the methods described above were performed using the \textsc{GammCor} code.~\cite{gammcor} Molpro has been also used to carry out NEVPT2\cite{nevpt2} calculations, which served as a reference for comparison with the low-order density-matrix methods.}}
The CASSCF and subsequent RDMs calculations were carried out either in Molpro~\cite{Molpro:12} or PySCF~\cite{pyscf2, pyscfrecent, pyscf} programs (see the Supporting Information for details). All post-CASSCF energy calculations were performed using the \textsc{GammCor} code.~\cite{gammcor} The NEVPT2\cite{nevpt2} calculations, which served as a reference for comparison with the low-order density-matrix methods, were performed out with Molpro, PySCF, or Dalton~\cite{dalton}, as described in the Supporting Information.

% For MC-srDFT, referred to below as ``srDFT'', the energy is computed according to Eqs.~(\ref{MC-DFT})--(\ref{EHxc}) using the short-range PBE exchange-correlation functional.\cite{goll2005short} MC-PDFT and MC-srPDFT energies, denoted ``PDFT'' and ``srPDFT'', respectively, are evaluated from the expressions given in Eqs.~(\ref{PDFT}) and (\ref{CASsrPDFT}). srPDFT employs the short-range PBE functional\cite{goll2005short} together with translated densities, see Eq.~(\ref{translation}). PDFT is based on the same translation scheme and employs the PBE functional,\cite{perdew1996generalized} see Eq.~(\ref{PDFT}). A standard value of the range-separation parameter, $\mu=0.4$ a.u., was employed in all range-separated integrals and functionals. Finally, the CAS-AC0 energy, denoted ``AC0'', is computed as described in Sec.~\ref{sec:AC}.

As representative multireference systems with low singlet--triplet (ST) gaps, we consider a set of organic biradicals from Ref.~\citenum{stoneburner2018mcp}, comprising cyclobutadiene and its derivatives, as well as the cyclopentadienyl cation. Geometries were taken from Ref.~\citenum{saito2011symmetry}. The standard active space employed in the CASSCF calculations includes all valence $\pi$ orbitals of carbon atoms belonging to or adjacent to the ring, as described in Ref.~\citenum{drwal2022efficient}. This choice is close to the $\pi$CPO scheme considered in Ref.~\citenum{stoneburner2018mcp}. Calculations were performed with the maug-cc-pVTZ basis set,\cite{papajak2009efficient} and the errors in the ST gaps were evaluated with respect to accurate DEA-EOMCC(4p--2h) reference data from Ref.~\citenum{stoneburner2017systematic}; see the Supporting Information for details.

Excited-state performance is assessed using singlet and triplet excitations compiled by Schreiber et al.\cite{Thiel:08} Geometries, active spaces, and basis sets were taken from that work. State-specific CASSCF calculations were carried out as described in Ref.~\citenum{tucholska2026mul}. The def2-TZVP basis set\cite{def2-tzvp} was employed, and errors were evaluated relative to CC3 excitation energies.\cite{Thiel:08,Schapiro:13,kannar2014benchmarking}

Spin-state energetics of transition-metal systems is illustrated on the example of three iron complexes: [Fe(H$_2$O)$_6$]$^{3+}$, [Fe(H$_2$O)$_6$]$^{2+}$, and [Fe(NH$_3$)$_6$]$^{2+}$. The geometries were taken from Refs.~\citenum{radon2019benchmarking} and \citenum{floser} for the first and the latter two complexes, respectively, and the spin states were adopted from the same references. The spin quantum numbers for the high-spin (HS) and low-spin (LS) states are S$_H=5/2$ and S$_L=3/2$ for the Fe(III) complex, and S$_H=2$ and S$_L=0$ for the Fe(II) complexes.
The computed spin-state splittings are defined as energy differences between the HS and LS states. The CASSCF wave function calculations are based on the standard protocol for transition-metal complexes established by Pierloot\cite{pierloot2003caspt2} and Roos et al. \cite{veryazov2011select}. This protocol involves  an active space comprising five Fe 3d orbitals, five Fe 4d orbitals, and two ligand orbitals responsible for $\sigma$ bonding with the Fe 3d shell, corresponding to (9e,12o) and (10e,12o) active spaces for the Fe(III) and Fe(II) complexes, respectively. The basis sets employed were cc-pwCVTZ-DK\cite{balabanov2006basis} for non-hydrogen atoms and cc-pVTZ-DK\cite{de2001parallel} for hydrogen atoms. The scalar-relativistic DKH2 Hamiltonian\cite{reiher2004exact} was used throughout.
Errors in the HS--LS energy differences were evaluated against benchmark values taken from Refs.~\citenum{radon2019benchmarking} and \citenum{domingo2010spin} for the Fe(III) and Fe(II) complexes, respectively; see the Supporting Information for details.
%
%Table~\ref{tab:Fe} presents the errors obtained from post-CASSCF calculations with dynamic correlation included via: spin-density srDFT with the short-range PBE exchange-correlation functional\cite{goll2005short}; srPDFT with the translated short-range PBE functional\cite{goll2005short}, see Eq.~(\ref{translation}); PDFT with the translated PBE functional\cite{perdew1996generalized}, see Eqs.~(\ref{PDFT}) and (\ref{translation}); and AC0 based on the ph-ERPA.

\subsection{Singlet-triplet gaps of  biradicals}
Singlet--triplet gaps for organic biradicals studied in Ref.~\citenum{stoneburner2018mcp} provide a demanding benchmark for multireference methods, as accurate ST splittings require a balanced treatment of static and dynamic correlation: the singlet states are strongly multireference, whereas the triplet states are typically dominated by a  single determinant.\cite{stoneburner2018mcp} Since the ST gaps range from about 0.1 to over 3 eV, Table~\ref{tab:biradicals} reports both the mean unsigned errors (MUE) and mean unsigned percentage errors (MU\%E).

% table biradicals
\begin{table*}
  \centering
  \caption{Mean unsigned errors (MUE),  mean unsigned percentage errors (MU\%E), and signed error distributions for the singlet-triplet gaps of organic biradicals from Ref.\citenum{stoneburner2018mcp}. Each thin line represents a single datapoint; the thick lines indicate the mean signed errors. Errors are given in eV. For computational details see Section~\ref{sec:3} and the Supporting Information.}
    \renewcommand{\arraystretch}{1.4}
    \setlength{\tabcolsep}{12pt}
     \sffamily
    \includegraphics{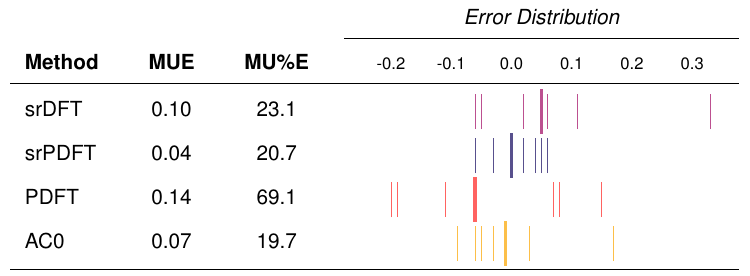}
    
    \vspace{1em}
    \footnotesize
    \label{tab:biradicals}
\end{table*}

 Both srDFT and srPDFT, which recover dynamic correlation via short-range functionals, perform well for the biradical ST gaps. 
They yield mean errors of 0.10 and 0.04~eV, respectively, corresponding to
percentage errors of approximately 20\%. Relative to CASSCF, this amounts to error reductions by factors of two and four (see Table~S1 in the Supporting Information). The good accuracy of srPDFT for biradical ST gaps is consistent with its satisfactory description of magnetic coupling in model non-Kekulé diradicals reported in Ref.~\citenum{rodrigues2023multi}.

%Interestingly, srDFT, which employs the charge density as a variable, increases the error to 0.50 eV (see SI). Replacing the variables in the spin functional with translated densities, as done in srPDFT, substantially reduces the error.

PDFT has previously been applied to biradicals by Stoneburner et al.\cite{stoneburner2018mcp} and more recently by Rodrigues et al.\cite{rodrigues2023multi}. Various translation schemes have been explored. The basic translation, see Eq.~(\ref{translation}), employed in the PBE functional leads to a mean error of 0.14~eV (Table~\ref{tab:biradicals}), only slightly larger than the 0.09 eV achieved with the fully translated PBE functional in Ref.~\citenum{stoneburner2018mcp}. This comparison is direct, since the latter calculations used identical basis sets and $\pi$CPO active spaces practically equivalent to those adopted here from Ref.~\citenum{drwal2022efficient}. 

In Ref.~\citenum{rodrigues2023multi}, a complex translation scheme was introduced, in which the translated densities are extended into the complex plane whenever the argument of the square root in Eq.~\eqref{translation} becomes negative. For  a set of 13 biradicals,  this modification reduced the mean error of the ST gaps  from 0.26~eV to 0.21~eV compared to the standard PBE translation when  CAS-$\pi$ active spaces were used. Note that the data set employed in Ref.~\citenum{rodrigues2023multi} was larger than the one  considered here. 

Without relying on DFT ingredients, ffAC0 (Table \ref{tab:biradicals} and Table S1 in the Supporting Information; see also Refs.~\citenum{tucholska2024dua} and \citenum{tucholska2026mul}, noting that a smaller basis set was used in those works) achieves an accuracy comparable to srPDFT, with a deviation of approximately 20\% relative to the coupled-cluster reference (MUE = 0.07~eV). The phAC0 variant is less accurate, exhibiting a mean error of 56\% (MUE = 0.15~eV).\cite{tucholska2026mul} These systems are also challenging for second-order correlation methods, with NEVPT2 and CASPT2 showing similar relative errors of about 50\% and 47\%, respectively.

% As demonstrated by Drwal et al.\cite{drwal2022efficient} and later by Tucholska and Pernal\cite{}, these systems are truly challenging for second-order correlation methods: the mean error of the phAC0 variant is as large as 56\% (MUE = 0.15~eV), while NEVPT2 and CASPT2 exhibit similar relative errors of 55\% and 47\%, respectively. For aminocyclobutadiene (C$_4$H$_3$NH$_2$) with an ST gap of only 0.12~eV (2.7~kcal/mol), both phAC0 and NEVPT2 even predict the wrong sign for the splitting.

\subsection{Excited states}

Vertical excitation energies are a standard way of assessing a method's ability to describe electronically excited states. Since its introduction in 2008, the dataset of Schreiber et al. \cite{Thiel:08,silva2010ben}, often referred to as Thiel's set, has become one of the most popular benchmarks, covering valence excitations in 28 representative closed-shell chromophores. Here, we first briefly review previous applications of MC-srDFT, MC-PDFT, and AC methods to excited states, focusing in particular on Thiel's set. We then compare these approaches directly for singlet and triplet transitions across the full set, and discuss the present results in relation to trends reported in the literature.

The first application of MC-srDFT to excited states employed a time-dependent extension of the method based on linear response (LR).\cite{fromager2013multi}
Using nucleobases\cite{Hubert:16b} and 32 singlet transitions from Thiel's set as a benchmark,\cite{hubert2016inv} LR-CAS-srDFT was shown to rival the accuracy of PT2 methods, with mean unsigned errors (MUEs) in the 0.3--0.4~eV range with respect to CC3 reference values. In these studies, srPBE \cite{goll2005short} was identified as the most reliable SR functional, and the $\mu=0.4$~a.u.\ value was recommended for the RS parameter. Using the full set, Hedeg\r{a}rd\cite{hedegaard2017ass} demonstrated that CAS-srPBE also affords oscillator strengths in good agreement with the CC2 results. Kjellgren et al.\cite{kjellgren2019tri} extended LR-MC-srDFT to triplet excitations and reported average deviations of 0.2--0.3~eV from CC3 results, depending on the functional, for selected systems from the sets of Thiel \cite{Thiel:08} and Loos\cite{loos2012mou}. Importantly, the generalized Tamm-Dancoff approximation was shown to be mandatory for reliable predictions for the triplet response. The same trends were observed for srDFT combined with generalized valence bond perfect-pairing wave functions.\cite{hapka2025time} 

Note that the standard srDFT choice $\mu=0.4$~a.u.\ recommended for excitation energies and oscillator strengths does not necessarily transfer to other properties. For example, satisfactory spin-spin coupling constants in transition metal complexes were obtained within LR-MC-srDFT only after increasing the range-separation parameter to $\mu=1.0$~a.u.\cite{kjellgren2021mul,hapka2025time} More recently, Hapka et al.\cite{hapka2026opt} proposed a tuning procedure for the range-separation parameter in srDFT. The resulting optimally tuned $\mu$ values typically fell in the 0.2--0.3~a.u. range and led to significantly more accurate molecular polarizabilities compared to the standard choice.

Although the LR treatment offers several advantages, such as accounting for orbital-relaxation effects, the state-averaged  and multi-state approaches remain the methods of choice for quasi-degenerate problems, which are commonly encountered in photochemical studies. In the non-variational, i.e., post-CASSCF approach, state-specific density matrices obtained from SA CASSCF calculations can be employed to evaluate the energy functional. This strategy has been commonly adopted in MC-PDFT. The promising performance of MC-PDFT demonstrated in early applications \cite{li2014multiconfiguration,hoyer2015mul}, including challenging intermolecular charge transfer excitations,\cite{ghosh2015mul} was later systematically assessed by Hoyer et al\cite{hoyer2016mul}, who investigated 23 low-lying excitations of various character in 18 molecules, with 13 systems overlapping with the Thiel's database. For this benchmark, translated PBE achieved an accuracy comparable to that of CASPT2, reaching MUE of approximately 0.2~eV. 

Recently, two independent variational formulations of srPDFT with state-averaging have been proposed.\cite{helmich2025exc,rodrigues2026state} Both studies benchmarked srPDFT against singlet excitation energies from  Thiel's set: Helmich-Paris et al.\cite{helmich2025exc} considered 139 data points using aug-cc-pVTZ, whereas Rodrigues et al\cite{rodrigues2026state} selected a subset of 27 low-lying transitions and employed the smaller aug-cc-pVDZ basis set. In spite of these differences, the authors showed that srPDFT consistently outperforms conventional PDFT. Ref.\citenum{helmich2025exc} compared complex-translated LDA and PBE, which gave MUEs of 0.68~eV and 0.65~eV, respectively, with their short-range variants, for which the MUE was reduced to 0.25 eV in both cases. 

In contrast to single-reference AC formulated within the DFT framework, multireference AC can be naturally applied for excited states.\cite{Pastorczak:18b,drwal2021exc} Although AC methods based on particle-hole ERPA perform on a par with PT2 for ground-state correlation energies,\cite{Pastorczak:18a,pastorczak2019capturing} they systematically overestimate singlet excitations.\cite{Pastorczak:18b,pastorczak2019capturing,hapka2020long} The pertinent deexcitation correction to the correlation energy proposed by Drwal et al.\cite{drwal2021exc}  alleviates this issue. The resulting AC0D significantly improved the excitation energies, reducing the MUE of AC0 from 0.8~eV to 0.4~eV, with respect to CC3 reference values for a subset of 74 singlet excitations from Thiel's set. Nevertheless, relatively large errors persisted for higher-lying excitations, making AC0D ultimately less accurate than NEVPT2, for which the corresponding MUE amounted to 0.3~eV.\cite{drwal2021exc} A subsequent study\cite{drwal2024mul} revealed a similar picture for triplet states: AC0D deviated on average by 0.3~eV for low-lying excitations and by 0.4~eV when the whole spectrum was considered. In the same work, Drwal et al.\cite{drwal2024mul} analyzed oscillator strengths for systems from Thiel's set. They showed that AC0D excitation energies combined with ERPA transition density matrices correct through first-order in the coupling parameter $\alpha$ offer accuracy competitive with CASPT2, reaching MUE of only 0.04 a.u.\ with respect to CC3.

%\kp{\sout{Recently, Tucholska and Pernal\cite{tucholska2024dua,tucholska2026mul} introduced a multireference AC formalism based on the particle-particle decomposition of the 2-RDM. The practical implementation of ppAC relies on the solution of ppERPA equations, which reduce in the single-reference limit to ppRPA, orignally developed by Yang and co-workers.\cite{van2013exchange,yang2013dou,yang2013ben} The ppAC method can be rigorously combined with phAC by exploiting the duality between the corresponding correlation energy amplitudes.\cite{tucholska2024dua} This strategy is motivated by the identification of the amplitude components underlying the deficiencies of phAC0 for excited states.\cite{guo2024spinless} } \emph{ta czesc zostala przeniesiona do sec.2.4}}
The recently developed  ffAC0 approach\cite{tucholska2024dua,tucholska2026mul}, which rigorously combines particle-hole and particle-particle AC0 approximations, performs comparably to PT2 methods. For the full Thiel set, both ffAC0 and NEVPT2 exhibit MUEs of approximately 0.3~eV for singlet and 0.2~eV for triplet transitions relative to the CC3 benchmark.\cite{tucholska2024dua} In contrast to AC0D,\cite{drwal2021exc,drwal2022efficient} the ffAC0 method remains accurate across the entire excitation spectrum.

We now compare srDFT (spin-functional), srPDFT, PDFT, and AC0 (ffAC0 variant) for singlet and triplet excitations from Thiel's set, comprising 140 and 71 transitions, respectively. All results were obtained with state specific (SS) CAS references (for SA-CAS-based data, see  Table S3, S4, and Figures S1 and S2 in the Supporting Information). The error statistics for singlet excitations, collected in Table~\ref{tab:ThielS}, reveal that srPDFT is the most accurate method in the group, with a MUE of 0.2~eV and a standard deviation (SD) of the same magnitude. This matches the accuracy achieved by variational srPDFT implementations employing complex-translated (ct) flavors of the SR functionals, as reported in Refs.\citenum{helmich2025exc} and \citenum{rodrigues2026state}. In fact, the ct variant of srPBE exhibits a visibly larger error spread, with an SD of 0.3~eV (see Table~3 in Ref.\citenum{helmich2025exc}). These results suggest that accounting for imaginary spin density contributions arising in Eq.~\eqref{translation} does not improve the numerical accuracy for this benchmark. Finally, the srPDFT error statistics closely resemble the performance of NEVPT2 based on the SA-CAS reference, see, e.g., Refs.\citenum{Schapiro:13} and \citenum{tucholska2026mul}.

Removing the OTPD from the SR functional, i.e., reverting to srDFT, nearly doubles the errors relative to srPDFT, with a MUE of 0.4~eV and an SD of 0.5~eV. For comparison, LR-MC-srDFT\cite{hedegaard2017ass} performs somewhat  better for the full singlet set (MUE of 0.3~eV),  and smaller error spread (SD = 0.4~eV). A further deterioration is observed for pure PDFT, where range separation is removed altogether. This method systematically underestimates singlet excitation energies, emerging as the least accurate of the low-order density-matrix approaches considered, with  MSE of $-0.5$~eV and both the MUE and SD above 0.6~eV. The variationally optimized ctPBE functional tested in Ref.\citenum{helmich2025exc} shows a practically identical MUE, but a markedly larger SD of 0.8~eV. 

The AC0 approach is the next most accurate method after srPDFT. AC0 systematically overestimates singlet excitations, which translates into a positive MSE of 0.2~eV. Although its MUE of 0.3~eV is larger than that of srPDFT, AC0 is the only other method with a narrow error distribution (SD = 0.3~eV), yielding a level of accuracy comparable to that of NEVPT2 with the same SS-CAS reference (MUE  = 0.3~eV and SD = 0.3~eV).\cite{tucholska2026mul}

The results for triplet excitations are summarized in Table~\ref{tab:ThielT}. The overall accuracy of the tested approaches is higher than that observed for singlet states, with MUEs falling in the 0.1--0.4~eV range compared to the 0.2--0.6~eV range for singlets (Table~\ref{tab:ThielS}). srPDFT remains the best-performing method, with both the MUE and SD equal to only 0.1~eV. This performance surpasses NEVPT2 and matches the CASPT2 accuracy\cite{Schapiro:13,tucholska2026mul} (see also Figure~\ref{fig:MUEs}). AC0 is again the second-best method. It retains a small error spread, with SD = 0.2~eV, but systematically overestimates triplet excitation energies, giving an MSE of 0.1~eV. 
In contrast to AC0, both PDFT and srDFT underestimate triplet excitation energies, achieving MSEs of $-0.1$ and $-0.3$~eV, respectively. Although PDFT still exhibits the largest error spread (SD = 0.4~eV), it is considerably more reliable for triplets than for singlets  and performs  better (MUE =0.3~eV) than srDFT (MUE =0.4~eV).

%table singlets Thiel
\begin{table*}
    \centering
    \caption{Statistical analysis and signed error distributions of 140 singlet excitation energies in the def2-TZVP\cite{def2-tzvp} basis set. Each thin line represents a single datapoint; the thick lines indicate the mean signed errors. All results were obtained using state-specific CASSCF references. Errors are reported with respect to the CC3 values.\cite{Thiel:08,Schapiro:13} The energy unit is eV. }
    %(srDFT: nie-spin, AC0: ff)
    \renewcommand{\arraystretch}{1.4}
    \setlength{\tabcolsep}{8pt}
    \sffamily
    \includegraphics{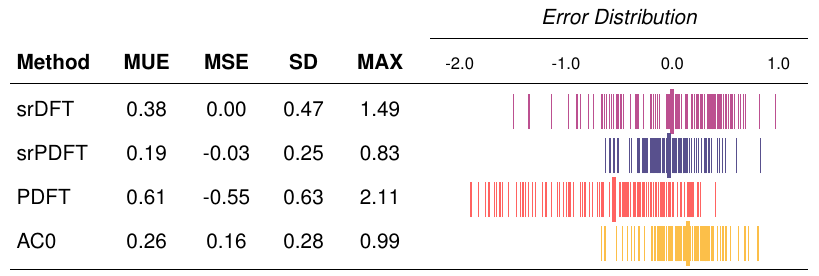}
\label{tab:ThielS}    
\end{table*}

%Thiel-triplet
\begin{table*}
    \centering
    \caption{Statistical analysis and signed error distributions of 71 triplet excitation energies in def2-TZVP\cite{def2-tzvp} basis set. Each line represents a single datapoint; the thick lines correspond to mean signed errors. All results using state-specific CASSCF references. Errors are reported with respect to the CC3 values.\cite{Thiel:08,Schapiro:13} The energy unit is eV. 
    %\ola{srDFT in the spin version}
    } %(srDFT: nie-spin, AC0: ff)
    \renewcommand{\arraystretch}{1.4}
    \setlength{\tabcolsep}{8pt}
    \sffamily
    \includegraphics{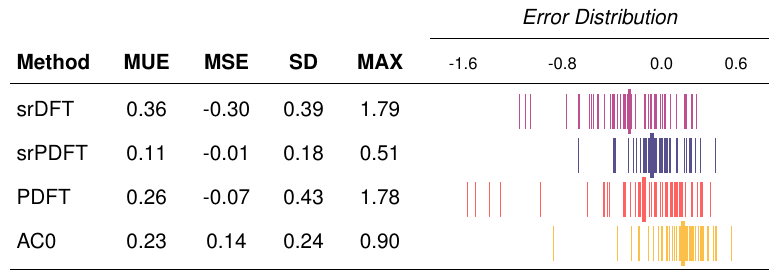}
\label{tab:ThielT}    
\end{table*}

\subsection{Spin states energetics of transition metal complexes}
\begin{figure*}
    \centering
    \includegraphics[width=\linewidth]{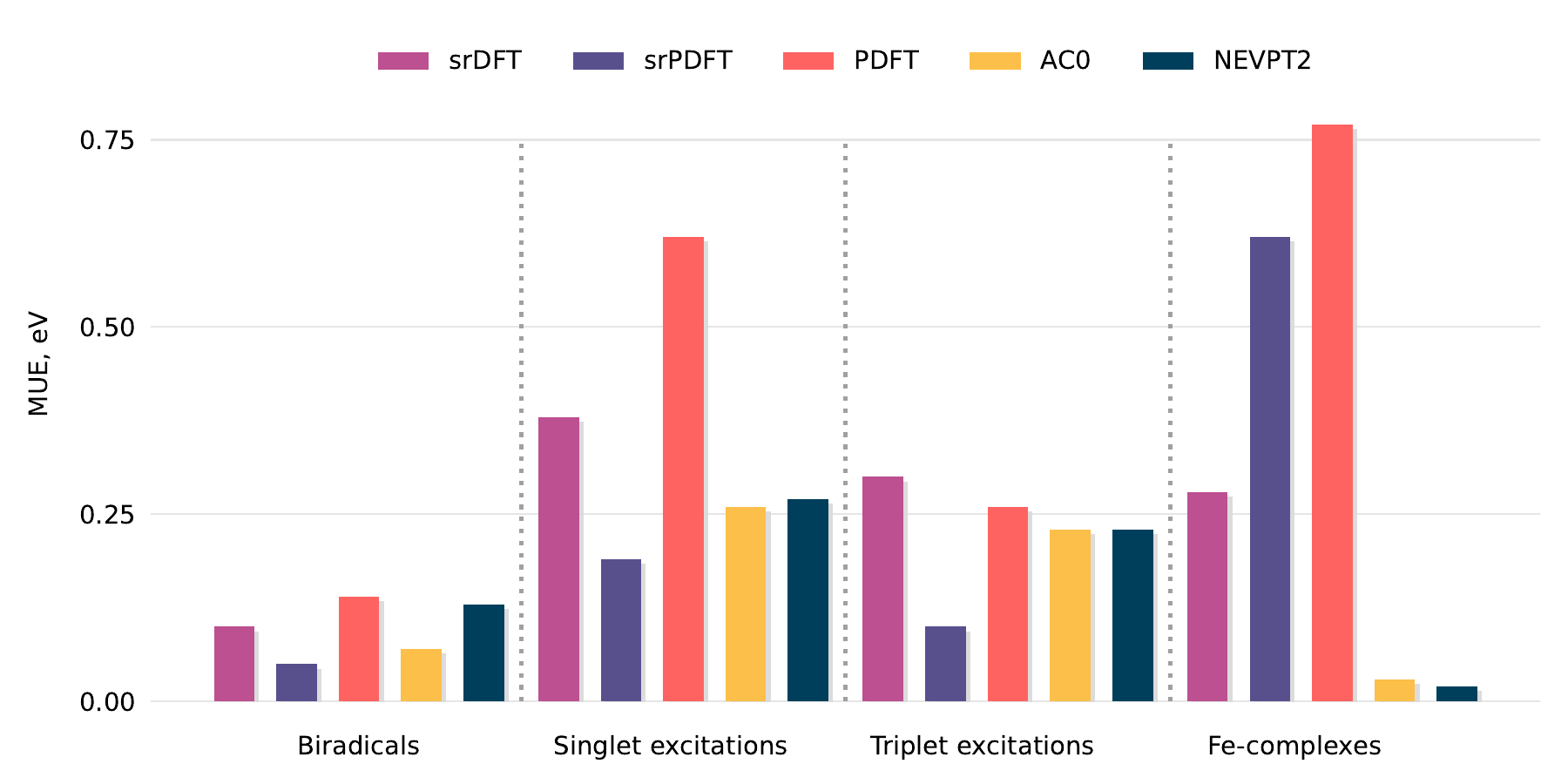}
    \caption{Mean unsigned errors across all data sets (see also Tables \ref{tab:biradicals}--\ref{tab:Fe}) for the low-order density-matrix methods for dynamic correlation, compared with NEVPT2.}
    \label{fig:MUEs}
\end{figure*}

Prediction of spin-state energetics in transition-metal complexes is one of the most demanding problems for electronic structure methods. As is well known, even a careful selection of active orbitals and the use of multireference wave functions are insufficient to achieve quantitative, or even qualitative, results without an adequate treatment of dynamic correlation energy. Incorporating the latter through density exchange-correlation functionals is particularly challenging. Indeed, a quick inspection of Table~\ref{tab:Fe} and Figure~\ref{fig:MUEs} shows that none of the DFT-based approximations considered here reproduces spin-state energetics with an accuracy competitive with that of \emph{ab initio} methods,  such as AC0 or NEVPT2.

Beginning with srDFT, Hedeg{\aa}rd et al.\cite{hedegaard2018mul} showed that extending the closed-shell formulation of the method to the open-shell case and employing SR spin-density functionals leads to reasonably accurate predictions of the spin-state splitting in the iron aqua complex [Fe(H$_2$O)$_6$]$^{3+}$, with errors of several tenths of an eV. This is consistent with the accuracy of approximately 0.3~eV obtained for the srDFT method reported in Table~\ref{tab:Fe}. The srDFT splittings in Table~\ref{tab:Fe} were computed using spin-density functionals (see Table S4 in the Supporting Information for the performance of the post-CASSCF srDFT variant employing a charge-density functional). Note that the HS--LS gaps in Table~\ref{tab:Fe} follow from post-CASSCF calculations, whereas those in Ref.~\citenum{hedegaard2018mul} were determined fully self-consistently. Apparently, the latter approach does not provide a noticeable improvement in accuracy over the post-CASSCF methodology.

\begin{table*}
    \centering
    \caption{Signed errors for the HS--LS gaps with respect to reference values (Acc.) of the iron complexes. The energy unit is eV.}
    \renewcommand{\arraystretch}{1.4}
    \setlength{\tabcolsep}{8pt}
    \sffamily
    \begin{tabular}{l c c c c c}
        \textbf{System} & \textbf{Acc.} & \textbf{srDFT} & \textbf{srPDFT} & \textbf{PDFT} &  \textbf{AC0}\\
        \cmidrule(lr){1-6}
        \ce{[Fe(H_2O)_6]^{3+}} & -2.06\cite{radon2019benchmarking} & 0.29 & 0.70 & 1.01 & 0.01 \\
        \ce{[Fe(H_2O)_6]^{2+}} & -1.45\cite{domingo2010spin}&  0.29 & 0.54 & 0.59  & 0.10\\
        \ce{[Fe(NH_3)_6]^{2+}} & -0.66\cite{domingo2010spin}&  0.25 & 0.62 & 0.70 & 0.00\\
        \midrule
    \end{tabular}
    \label{tab:Fe}
\end{table*}

Unlike in the applications discussed earlier, replacing the spin-density arguments in the SR exchange-correlation functional with translated densities, as done in srPDFT, does not improve the accuracy of srDFT. On the contrary, srPDFT roughly doubles the error in spin-state energetics relative to srDFT, cf.\ Table \ref{tab:Fe}. This observation is consistent with the poor performance of srPDFT reported recently in Ref.~\citenum{helmich2025exc}, where various srPDFT methods did not provide a consistent improvement over CASSCF excitation energies for same-spin excited states of  Co(II), Fe(II), and Ni(II) aqua complexes. Even worse, all investigated variants failed to predict the correct state ordering for all systems.

A closer inspection of the results for the iron aqua and ammonia complexes reveals that, in all cases, uncorrected CASSCF overestimates the HS--LS energy gaps (see Table S6 in the Supporting Information). Adding the srPDFT energy correction with the range-separation parameter $\mu=0.4$~a.u.\ changes the gaps in the expected direction: the LS states are stabilized relative to the HS states and the gaps decrease. However, this stabilization is too strong, leading to an underestimation of the HS--LS splittings. Thus, it is not surprising that decreasing $\mu$ from 0.4 to 0, which corresponds to the PDFT limit, further overstabilizes the LS states.  Consequently, PDFT performs worse than srPDFT, with the MUE for the HS--LS gaps approaching 0.8~eV (see Table~\ref{tab:Fe} and Figure~\ref{fig:MUEs}).

Previous applications of PDFT to Fe(II) and Fe(III) complexes with ligands spanning from weak- to strong-field regimes have shown that the method is only qualitatively correct, reproducing trends in spin-state splittings obtained with more expensive perturbative approaches.\cite{wilbraham2017mul} In particular, PDFT significantly underestimates the HS--LS gaps for weak-field ligands such as H$_2$O and NH$_3$, while overestimating them for strong-field ligands. Similar conclusions were reached in studies of iron complexes with larger ligands, for which average absolute errors of approximately 0.7~eV were reported for the tPBE functional, although some reduction of the errors was achieved using specially tailored functional modifications.\cite{stoneburner2020transition}

% \mh{\sout{Turning to the performance of the multireference adiabatic connection method AC0 (based on phERPA) for the three iron complexes included in the discussion, it is excellent} 
The performance of the multireference AC0 method based on phERPA is excellent for the studied test set: for each system, the magnitude of the HS–LS gap agrees with the benchmark to within 0.1 eV. The results are practically identical to those of NEVPT2, see Figure~\ref{fig:MUEs} and Table S6 in the Supporting Information. 
In Ref.~\citenum{guo2024spinless},
it was demonstrated that the AC0 and NEVPT2 equations are indeed similar in selected perturbation subspaces. 
% \mh{\sout{In those subspaces that are most expensive for NEVPT2, i.e., those requiring 3- and 4-electron RDMs, the two methods differ} 
The methods differ in the subspaces that are most expensive for NEVPT2, whose treatment requires 3- and 4-electron RDMs: NEVPT2 is based on doubly excited perturbers, whereas AC0 approximates the corresponding contributions by coupling two single excitations.~\cite{guo2024spinless} The numerical closeness of AC0 and NEVPT2 for iron complexes is stunning, especially considering  that AC0 relies only on 1- and 2-RDMs.

Our results for iron complexes are in line with those in Ref.~\citenum{tzeli2024importance}, where calculations on $[2\mathrm{Fe}-2\mathrm{S}]$ model systems with Fe(II) and Fe(III) oxidation states showed that AC0 is as accurate as NEVPT2. Similarly, an earlier study of iron--porphyrin systems and spin states of iron--sulfur complexes established AC0 as a reliable tool for studying the energetics of transition metal compounds.~\cite{beran2021density}

We conclude that for spin-state energetics of transition metal complexes, only AC0 delivers consistent and reliable performance, providing spin-state energy splittings with an accuracy comparable to that of more expensive PT2 methods. In contrast, the discussed DFT-based approaches are significantly less accurate and may exhibit erratic behavior.

\section{Summary and Perspective}\label{sec:4}
In this Perspective, we presented methods capable of recovering dynamic correlation energy for multireference wavefunctions exclusively from low-order reduced density matrices and densities, using at most the 2-RDM of the reference wavefunction. The reviewed and benchmarked approaches possess a number of desirable features: (i) universal applicability, in particular to electronic states of arbitrary spin symmetry while respecting the spin symmetry imposed on the reference wavefunction, (ii) preservation of size consistency of the underlying wavefunction, and (iii) essentially nonempirical character. Two classes of methods were considered: approaches incorporating DFT ingredients, i.e.\ modified exchange-correlation functionals, and \emph{ab initio} multireference adiabatic connection methods.

The DFT-based approaches comprise srDFT, which recovers dynamic correlation through a short-range functional depending on the electron density, as well as PDFT and srPDFT, which employ ``translated'' exchange-correlation functionals  incorporating  the on-top pair density.  Within the adiabatic connection framework, we focus on AC0 approximations, which rely on the 1- and 2-RDMs of the reference wavefunction.

Although these methods exist in a number of variants, we discuss the most robust and extensively tested formulations. All calculations presented here  were performed within a post-CASSCF framework. To ensure a reliable benchmark, we adopted computational setups regarded as standard for each problem, including the choice of CASSCF active spaces and basis sets. The selected benchmark data sets represent challenging multireference problems, and the obtained results were analyzed in the broader context of previously reported literature data for related systems and alternative variants of the methods.

A primiary advantage of low-order-RDM methods over standard multireference dynamic-correlation approaches based on CI, CC, PT2, and related formalisms,\cite{cheng2022post} is their favorable computational efficiency. Within a post-CASSCF framework, srDFT is the least computationally demanding since it depends only on the one-electron density. Its additional overhead arises from the need to evaluate Coulomb integrals involving the modified long-range electron interaction alongside the standard  ones. For PDFT and srPDFT, the bottleneck is the construction of the on-top pair density from the 2-RDM on the DFT grid. This step scales as the fourth power of the number of active orbitals multiplied by the number of grid points. In the case of AC0, the overall scaling is fifth order with respect to the system size, while the active-space-dependent steps scale as the sixth power of the number of active orbitals.

The relatively modest computational cost makes low-order-RDM methods attractive for applications involving large active spaces ($>$ 50 active orbitals). This possibility has already been successfully demonstrated for AC0.~\cite{zuzak2024surface}  Moreover, AC0 has a well-defined FCI limit: upon systematic enlargement of the active space, the AC0 correction vanishes and the total energy converges toward the FCI result.

Applications of the DFT-based approaches to large active spaces require greater caution. Although srDFT is, in principle, free from correlation double counting, the approximate nature of the SR functionals implies that enlarging the active space does not necessarily improve the results. Owing to the range separation of the electron interaction, the total energy in MC-srDFT calculations  saturates rapidly with increasing active-space size,\cite{Ferte:19} so further active-space expansion mainly increases the computational cost,  without improving accuracy. For PDFT and srPDFT, the situation is potentially more problematic, since these methods are not rigorously free from correlation double counting. Consequently, extending the active space to very large sizes could even deteriorate the results.

An appealing feature of srDFT-based approaches is their faster convergence with respect to the basis-set size compared with fully \emph{ab initio} correlation methods. This stems from the fact that the short-range correlation functional relies only on the electron density. In contrast, functionals incorporating the on-top pair density, while still based on local quantities, exhibit a more pronounced basis-set dependence. Because the OTPD probes electron correlation at the electron-coalescence point, it is particularly sensitive to the quality of the basis set. Nevertheless,  the basis-set dependence of srDFT remains weaker than that of PT2 methods.\cite{king2022lar}
Although the AC0 correlation energy converges slowly with basis-set size, as is typical of \textit{ab initio} methods, recently proposed basis-set incompleteness corrections\cite{giner2018curing,hapka2025correcting} can be applied at marginal additional cost, providing an effective route toward the complete-basis-set limit.

Comparison of the mean unsigned errors of low-order-RDM methods for four selected data sets (Figure \ref{fig:MUEs}) demonstrates that both the DFT-based and AC approaches are capable of achieving the desired accuracy.
The results clearly indicate that the use of short-range functionals is advantageous: on average, srDFT and srPDFT perform better than PDFT. In particular, srPDFT, which employs translated short-range functionals, exhibits  excellent performance for nearly all applications considered here. A notable exception concerns iron clusters, for which  srDFT approaches depending explicitly on the physical spin density yield more accurate results. AC0  shows consistently high accuracy across all benchmark sets, with errors comparable to those obtained with the perturbative NEVPT2 method. Overall, it appears to be the most robust and reliable among the benchmarked low-order RDM methods.

%Although our focused here is on energetics, other properties can also be obtained using low-rank-RDM appoaches within short-range-DFT-based and AC methods. In the former case, oscillator strengths can be computed with good accuracy,\cite{hedegaard2017ass} AC-ERPA-based corrections to CASSCF oscillator strengths have also been proposed.\cite{drwal2024mul} More recently it has been also shown that tuning the range-separation parameter in the range-separated functional yields accurate molecular polarizabilities.\cite{hapka2026opt}

Ongoing efforts are devoted to further improving the accuracy of the discussed frameworks. For AC0, promising developments include relaxing the constraint of a fixed 1-RDM along the adiabatic-connection path,\cite{matouvsek2023toward} as well as PT2-motivated modifications of the treatment of single excitations.\cite{guo2025approximation} The latter improves correlation energies, although at the cost of introducing dependence on the 3-RDM. Less systematic but physically motivated corrections that avoid   higher-order RDMs have also been proposed, for example through short-range correlation corrections tailored for active-space wavefunctions .\cite{hapka2024self}

In DFT-based approaches, further gains in accuracy may arise from improved functional design, including the incorporation of known analytical constraints.\cite{Ferte:19} Modifications of the translation scheme, such as complex translation, have also been reported to improve the accuracy in selected applications.\cite{rodrigues2023multi}

The developments reviewed in this Perspective indicate that low-order-RDM-based approaches constitute a promising and rapidly evolving direction for multireference electronic-structure theory. In particular, methods depending only on the one- and two-electron reduced density matrices appear especially attractive in the context of quantum computing. Within hybrid quantum-classical schemes
%, such as those based on the variational quantum eigensolver (VQE),\cite{peruzzo2014var} 
the multireference wavefunction and the corresponding low-order RDMs can be generated on a quantum device, while the remaining dynamic correlation energy may be recovered efficiently on a classical computer. The promise of such approaches has recently been demonstrated in the context of PDFT\cite{boyn2021quantum,chen2026quantum} and AC0.\cite{matousek2024variational}

\section{Supporting Information}
The Supplementary Information provides complete computational details, active space definitions, and optimized Cartesian coordinates for the studied biradicals, organic chromophores, and iron complexes. It includes tabulated total energies, singlet-triplet gaps, high-spin-low-spin energy splittings, natural orbital occupation numbers, and orbital isosurface visualizations.  For the Thiel set, separate spreadsheet files (.xlsx) are provided containing absolute energies, vertical excitation energies, and computed errors relative to reference values for the complete set of singlet and triplet transitions. Additionally, for the Thiel set, we present data obtained from self-consistent-MC-srDFT calculations using both the short-range PBE and LDA functional variants. For triplet excitations, results from post-CASSCF spin-srDFT calculations are also provided, where the short-range exchange-correlation functional explicitly incorporates a dependence on the physical spin density rather than the charge density alone. Full details regarding the software packages and codes used to generate the reference wave functions and compute the dynamic correlation corrections are explicitly documented.

\begin{acknowledgement}
The National Science Center of Poland supported this work under grants no.\ 2021/43/I/ST4/02250 and no.\ 2021/43/D/ST4/02762. For the purpose of Open Access, the author has applied a CC-BY public copyright licence to any Author Accepted Manuscript (AAM) version arising from this submission.
\end{acknowledgement}

%\vspace{0.5cm}
%\section{Data Availability}
%The data that support the findings of this study are available within the article and its supplementary material. The raw data are available in the Zenodo repository at 10.5281/zenodo.XXXX.

\vspace{0.5cm}

\bibliography{biblio}

\end{document}